\documentclass{article}
\usepackage{lmodern}

\usepackage{graphicx} 
\usepackage{authblk}
\usepackage{subcaption} 
\usepackage{wrapfig}
\usepackage[table]{xcolor}
\usepackage{colortbl}
\usepackage{mathtools}
\usepackage{amsmath}
\usepackage{hyperref}
\usepackage{array} 
\usepackage{float}
\usepackage{multirow}
\usepackage{tikz}
\usetikzlibrary{positioning, matrix, arrows, calc}
\usepackage{comment}
\usepackage{etoolbox} 
\DeclareGraphicsExtensions{.pdf,.png,.jpg}

\author{Adeela Bashir$^1$}
\author{Zia Ush Shamszaman$^{1,2}$}
\author{Zhao Song$^1$}
\author{The Anh Han$^{1,2,*}$}
\affil{$^{1}$School of Computing, Engineering and Digital Technologies, Teesside University\\ $^{2}$Center for Digital Innovation, Teesside University\\ $^*$Corresponding: The Anh Han (T.Han@tees.ac.uk)}

\begin{document}

\title{Co-evolutionary Dynamics of Attack and Defence in Cybersecurity} 

\date{\today}
\maketitle

\newcommand{\highlight}[1]{%
  \fcolorbox{red}{lightgray}{#1}
}

\begin{abstract}

In the evolving digital landscape, it is crucial to study the dynamics of cyberattacks and defences. This study uses an Evolutionary Game Theory (EGT) framework to investigate the evolutionary dynamics of attacks and defences in cyberspace. We develop a two-population asymmetric game between attacker and defender to capture the essential factors of costs, potential benefits, and the probability of successful defences. Through mathematical analysis and numerical simulations, we find that systems with high defence intensities show stability with minimal attack frequencies, whereas low-defence environments show instability, and are vulnerable to attacks. Furthermore, we find five equilibria, where the strategy pair ”always defend and attack” emerged as the most likely stable state as cyber domain is characterised by a continuous battle between defenders and attackers. Our theoretical findings align with real-world data from past cyber incidents, demonstrating the interdisciplinary impact, such as fraud detection, risk management and cybersecurity decision-making.
Overall, our analysis suggests that adaptive cybersecurity strategies based on EGT can improve resource allocation, enhance system resilience, and reduce the overall risk of cyberattacks. By incorporating real-world data, this study demonstrates the applicability of EGT in addressing the evolving nature of cyber threats and the need for secure digital ecosystems through strategic planning and proactive defence measures. 
\end{abstract} 
\vspace{1cm}
\noindent\textbf{Keywords: }{\textit{Evolutionary Game Theory, Cyber Security Decision Support, Ransomware, Adaptive Defence Strategies, AI-Driven Threat Analysis, Behavioural Analysis, Cyber Attacks}
\section{Introduction}
In today's digital world, the frequency of cyberattacks is escalating, driven by technological advancements, use of Internet of Things (IoT) devices, cloud computing, and inadequate security practices. This expanding attack surface increases the vulnerability of interconnected systems, particularly those embedded with sensitive data in the Industrial Internet of Things (IIoT) within industry4.0 \cite{ahmad2024enhancing}. According to an estimate, cybercrime is projected to cost \$10.5 trillion annually by 2025,as attackers target valuable information such as Personally Identifiable Information (PII), Intellectual Property (IP), financial records, and operational data \cite{sharif2022literature} .

Cybercriminals use a variety of tactics such as malware infections \cite{jang2014survey}, ransomware attacks, Distributed Denial of Service (DDoS) assaults \cite{kumar2024effective}, phishing schemes, and more sophisticated methods like Advanced Persistent Threats (APTs) \cite{7892931} and zero-day exploits \cite{tariq2023critical}. Numerous high-profile ransomware incidents have underscored the urgency of addressing these threats \cite{o2018evolution}, with notable recent cases including {MOVEit} (2023), BlackCat/ALPHV (2021), REvil/Sodinokibi (2019-2021), DoppelPaymer (2019), Ryuk (2018), and {WannaCry} (2017) \cite{aljaidi2022nhs}, \cite{akbanov2019wannacry}. Critical sectors like government agencies, financial institutions, healthcare systems, and e-commerce platforms are compelled to defend their systems against these attacks to mitigate reputational and financial losses.

In response to the evolving threat landscape, researchers are exploring advanced attack detection methods leveraging AI-enabled attack detection methods \cite{jia2023artificial}, anomaly-detection \cite{nguyen2024deep} and deep learning methods \cite{rasikha2024ensemble}, and graph learning methods \cite{zhao2022cyber}. Businesses are more likely to use the Incident Response Planning (IRP) method to address cybersecurity challenges. IRP activates defensive mechanisms like firewalls, intrusion detection/prevention systems (IDS/IPS), encryption,honeypots and deception, risk management, and continuous monitoring of the system after a security breach \cite{yang2023highly}. While these technical solutions focus on detecting and resolving the attacks, understanding the attack behaviour is crucial to prevent and reduce the number of successful breaches.
\begin{figure}[H]
    \centering
    \includegraphics[scale=0.65]{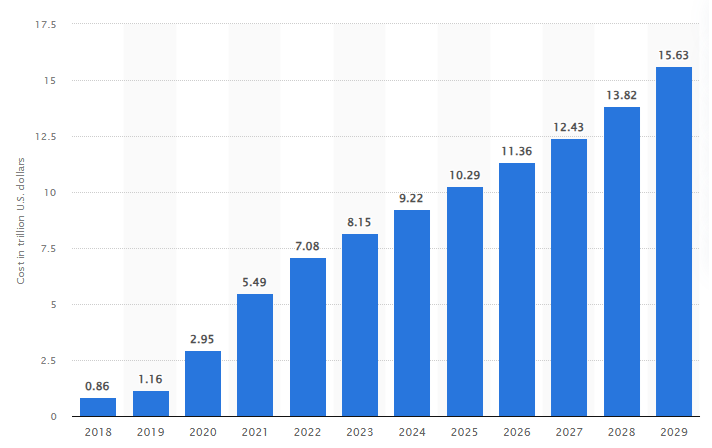}
    \caption{Estimated cost of cybercrime worldwide 2018-2029 (source: Statista Market Insights),    published by Ani Petrosyan, July, 2024}
    \label{fig:1}
\end{figure}
According to global indicator forecasts \cite{petrosyan2023estimated}, \cite{baruwal2024towards}, cybercrime costs are expected to rise significantly between 2024-2029, as shown in Figure \ref{fig:1}. This is due to the presence of several challenges that complicate effective cyber defence. Firstly, attackers constantly develop new tactics, techniques, and procedures (TTPs), outpacing traditional defence techniques. APTs and zero-day vulnerabilities are the biggest threats to cyber security due to their stealthy nature \cite{juuso2013proactive}. Secondly, many organizations, especially Small and Medium-sized Enterprises (SMEs), lack the financial and technical resources to adopt comprehensive cybersecurity solutions, making them susceptible to cyber-attacks \cite{heidt2019investigating, 9853515}. Vulnerabilities like human error, phishing attacks, sharing credentials, and malware installation increase vulnerabilities and threaten the cybersecurity in critical sectors such as healthcare, energy, and finance \cite{ilca2023enhancing}.

Given the increasing frequency of cyber threats targeting critical systems, it is imperative to investigate attack and defence patterns through system behavioural analysis \cite{skopik2022detecting, dorsey2003mathematical}. The dynamic interplay between attackers and defenders in cybersecurity can be analysed using Evolutionary Game Theory (EGT) \cite{smith1973logic}. EGT, an application of game theory to dynamical systems,  models continuous interactions, adaptation, and strategy evolution among game players   \cite{ohtsuki2008evolutionary,sigmund2010calculus}. EGT studies the evolution of strategies over time within a population based on their reproductive success, by considering bounded rationality and dynamic nature of the player interactions \cite{khalid2023recent, ohtsuki2008evolutionary}. EGT is crucial for stability analysis  \cite{sigmund2010calculus}, as its frequency dependent nature allows us to study how well a strategy performs compared to other strategies co-present in a population \cite{SMITH198643,hofbauer1998evolutionary}. 

The remainder of this paper is structured as follows: Section 2 discusses the motivation behind this research work. Section 3 provides an overview of the previous research on cyber defence using EGT. Section 4 presents, the cyber-attack defence game model and stability at equilibrium points analysis. Section 6 analyses the social welfare and parameters impact on it. Section 6 discusses the implications of these findings to address real-world cyber threats and concludes the paper with key insights.

\section{Research motivation}
Understanding player's behaviour in the cyber-attack and defence game is crucial, because it depends on the expected outcomes that are influenced by the opponent's actions \cite{anderson2006economics}. Sometimes, defenders may mistakenly think that their data is not valuable enough to attract attackers, yet attackers attempt an attack not just for financial benefits but for personal satisfaction \cite{patterson2019behavioral}. This sense of achievement often triggers small-scale attacks, where attackers exploit unsecured systems because attacks are less costly, less risky, and convenient \cite{von2013information}. Conversely, when defenders invest in robust defensive resources, attackers face increased difficulty and cost, thus reducing the likelihood of successful breaches. 

Based on these reasons, continuous improvement of defence mechanisms is essential in today's digital environment. However, deterring attacks requires more than just technological solutions, it requires a strategic approach that incorporates strategic modelling and behavioural analysis of the players \cite{tirenin1999concept}. The ability of EGT, to model strategic interactions and dynamic behaviours within populations makes it suitable for cybersecurity applications. EGT has been successfully applied in many fields---such as  computer science, physics, biology and economics \cite{hilbe2023evolutionary,Han2022emergent,perc2017statistical,traulsen2023future,cimpeanu2021cost,liu2020evolutionary,JIA2024111962,ZHU2025113153}---as well as diverse important application domains--- such as healthcare \cite{zhu2018evolutionary,alalawi2019pathways}, AI safety and governance \cite{cimpeanu2022artificial,han2020regulate}, and climate change mitigation \cite{milinski2008collective,santos2011risk}---creating  flexible solutions in a complex dynamics context.

In our study, we employed EGT to analyse the system dynamics of cyber attack and defence. By randomly generating extensive number of games, we aim to identify which strategies dominate within large populations. Rather than focusing solely on attacks and attackers only, we have discussed the scenario from defender's perspective, recognising that while attacks and attackers can't be controlled but defence can be improved by various factors. According to the NIST Cybersecurity Framework \cite{pascoe2023public}, defence process is divided into five phases, each with associated costs. Our model focuses on the most crucial phases of identifying and protecting assets, because here strategic decisions can enhance security before an attack happens.
\begin{figure}[H]
    \centering
    \includegraphics[scale=0.6]{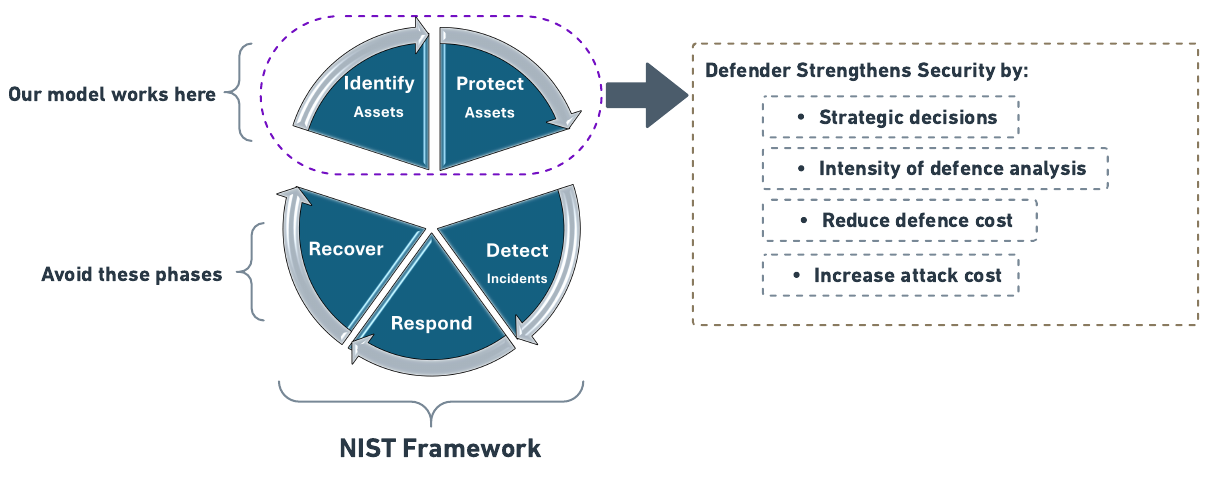}
    \caption{Our model assists in strategic decisions on the first two phases of NIST framework to mitigate the attacks by strengthening the defence.}
    \label{fig:2}
\end{figure}
As shown in the Figure \ref{fig:2}, 
by proactively identifying and protecting assets through robust protective measures, including latest AI technologies, defenders can reduce defence cost, increase attack cost, and avoid reactive phases---such as detection, response, and recovery---which are more costly. EGT helps organisations to allocate resources 
efficiently and stay ahead of evolving cyber-threats. Our model focuses on early-stage security enhancements, making defences cost-effective and helping cybersecurity decision-makers in strategic planning. 

Furthermore, we use EGT as a knowledge-based framework to model the complex interactions between attackers and defenders and how their strategies co-evolve over time within cybersecurity systems. This research bridges the gap between theoretical models and real-world applications, leveraging knowledge-based systems for strategic decision-making in cyber defence. In real-world scenarios, the probability of catching the attackers and imposing penalties is low because attackers can exploit the data across borders. Gaining benefits without any fear is the main reason behind the increase in cyber-attacks. To deter attackers, defenders must use strategies that increase the cost and difficulty of attacks by safeguarding sensitive data \cite{fan2022survey}.

This research motivation leads to the following hypotheses.

\textit{\textbf{H1:}} Attackers try to exploit systems with weaker security and higher benefits.

\textit{\textbf{H2:}} Defenders are more likely to invest in low-cost defences that target the high-impact attacks on confidential data.

\textit{\textbf{H3:}} Successful defence is achieved when the cost of defence is lower than its benefit while the cost of attack is higher than its benefit and the intensity of defence is increased.

We study evolutionary dynamics  between the populations of attackers and defenders by varying key factors including the attack and defence costs and probability of successful defence, to explore stable strategies and cybersecurity system dynamics. Our findings show that a high probability of defence is the most impactful factor in maintaining system stability. Additionally, keeping the defence cost low is crucial to motivate the defender to invest in system security. and at the same time high probability of defence should elevate the attack cost to discourage the attackers.

The key contributions of this work are:
\begin{enumerate}
    \item We developed a game-theoretic model using EGT to study evolutionary dynamics of cyber attack and defence, and analysed the stability of equilibrium points.
    \item Real-world cyberattack data is used to validate model parameters.
    \item We investigated social welfare and parameter impacts.
    \item We studied the effect of increasing probabilities of catching the attacker and imposing penalties.
\end{enumerate}

\section{Related Work}
The application of EGT to cyber security is a rapidly evolving field We review earlier studies that examined the application of game theory and EGT to cybersecurity issues. This highlights the theoretical developments and real-world applications of game theory that influence modern defence tactics like static and dynamic game models \cite{etesami2019dynamic}, stochastic games \cite{miao2018hybrid}, differential games for continuous decision-making \cite{zhang2023security}, replicator dynamics for strategy evolution \cite{boudko2019adaptive}, and Stackelberg games for leader-follower scenarios \cite{zhang2021bayesian}. Every method has its pros and cons in decision-making in cyber cybersecurity space.

The idea catching and punishing the attacker is proposed in \cite{yang2021multi} using a multi-player evolutionary game model to analyse the interactions between a defender and multiple attackers in a network security scenario. The long-term dynamics of the game and the evolution of strategies are observed over time, and two different punishment schemes (static and dynamic) are proposed depending on the attack intensity. Also, the importance of making laws and regulations to strengthen defence is discussed. However, the effectiveness of this scheme depends on the ability to catch attackers, which is a significant challenge in the cyber world, as most cyber attackers remain anonymous and are never caught.

Another method of formulating defence mechanisms is to plot various possible sets of behaviours and apply defence accordingly. For instance, in \cite{hu2020optimal} bounded rationality is used, and the authors proposed a stochastic evolutionary game model introducing a flexible parameter $(\lambda)$ in LQRD model to quantify the degree of rationality of players. The model categorizes players into different types, including  moderate, aggressive, and dangerous, which may not fully capture the diversity of behaviours exhibited by real attackers and defenders. In reality, it is not possible to make defensive strategies that are dependent on the types of attackers; as this trait is not always observable or measurable. 

To address the inherited issue of randomness in attack-defence process, stochastic games are used in \cite{xu2020study} and the strategy evolution of the players is analysed. The introduction of stochasticity promotes the defenders' ability to select optimal strategies but the effectiveness of the model relies on the accurate estimation of certain parameters, e.g., punishment to attackers and rewards for defenders, etc., which is challenging to determine in practice. To perform real-time analysis of defence strategy selection in rapidly changing environments, a differential game model for continuous decision-making is proposed in \cite{zhang2018attack}. Their model identifies optimal attack and defence strategies for the players but the real-time calculation of the optimal strategies, especially solving Hamilton-Jacobi-Bellman (HJB) equations for continuous systems, is computationally intensive and can overload the system in high-frequency attack environments. 

All the above studies focus on the attacker, emphasizing the importance of catching and penalising the attacker to reduce incidents. It is crucial to recognise that attackers cannot be entirely controlled from the defender's side. Rather, defenders should adopt strategies to strengthen their defences. For instance, smart home users rely on smart devices for their convenience, but to ensure safety they must remain vigilant about potential cyber threats and understand how to mitigate them\cite{app13074645}. This study highlights the importance of affordable training costs and rewards to encourage user participation, which can lead to a decrease in successful attacks. The study also emphasized the need of government support and national cybersecurity plans, to promote public awareness. However, beyond this theoretical analysis, simulations and analysis of real-world data is needed to better understand user behaviour and smart-home security strategies.

All techniques discussed so far involve the effort of the defender alone, but to promote a safer community and to secure national security infrastructure, a collaborative approach is proposed in \cite{TOSH201827}. The authors propose CYBer security information EXchange (CYBEX) framework to facilitate information sharing among organizations about breaches and fixes to better combat cyber attacks. EGT is used to investigate evolutionary dynamics within CYBEX to analyse conditions under which firms join and achieve stable strategies. Also, CYBEX adjusts participation costs to encourage cooperation and increase its own revenue.

Despite the increase in EGT applications to model cyberattacks and defences, there remains a gap in understanding crucial factors such as the attacker's adaptive learning process, probabilistic defence success rate, and the cost of implementing the defensive measures. These factors are critical for accurately modelling real-world scenarios, as attackers learn from failed attempts and adjust their strategies accordingly. Consequently, defenders must allocate their limited resources efficiently to strengthen their defences and counteract evolving threats. 
\section{Cyber Game Model}
In this model, we consider two populations: attackers and defenders, consisting of bounded rational players and each player has two strategies. In the attacker population, the strategies are "attack" and "no attack", while in the defence population, strategies are "defence" and "no defence", and the players interact between populations. 
Numerically, no attack and no defence strategies are denoted by 0, and attack and defence strategies are denoted by 1. 
Figure \ref{fig:3} illustrates  the model and interactions. 

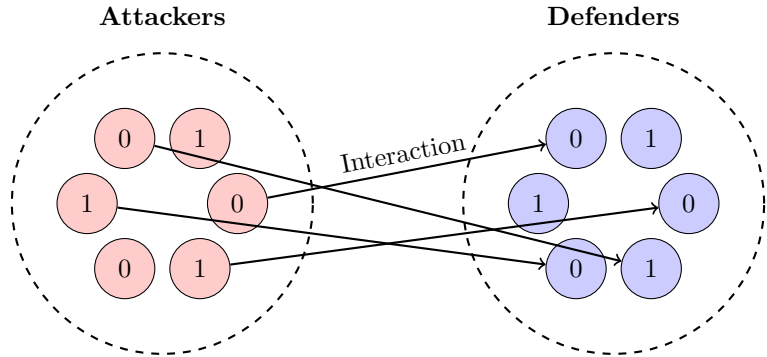
\begin{figure}[H]
    \centering
    \begin{tikzpicture}

        \draw[thick, dashed] (-3, 0) circle (2); 
        \draw[thick, dashed] (3, 0) circle (2);  

        \node at (-3, 2.5) {\textbf{Attackers}};
        \node at (3, 2.5) {\textbf{Defenders}};

        \foreach \i in {0, 60, 120, 180, 240, 300} {
            \node[draw, circle, fill=red!20, minimum size=0.8cm] (att\i) 
            at ({-3 + cos(\i)},{sin(\i)}) {\pgfmathparse{int(mod(\i/60,2))}\pgfmathresult};
        }

        \foreach \i in {0, 60, 120, 180, 240, 300} {
            \node[draw, circle, fill=blue!20, minimum size=0.8cm] (def\i) 
            at ({3 + cos(\i)},{sin(\i)}) {\pgfmathparse{int(mod(\i/60,2))}\pgfmathresult};
        }

        \draw[->, thick] (att0) -- (def120) node[midway, above, sloped] {Interaction};
        \draw[->, thick] (att180) -- (def240) node[midway, below, sloped] {};
        \draw[->, thick] (att300) -- (def0) node[midway, below, sloped] {};
        \draw[->, thick] (att120) -- (def300) node[midway, above, sloped] {};

    \end{tikzpicture}
        \caption{Cyber-attack and defence model. Red and blue circles  represent players in the attack and  defend  populations. No Attack and No Defence strategies are denoted by 0, and Attack and Defence strategies by 1.}
    \label{fig:3}
\end{figure}

The players are non-cooperative because they act independently, and their bounded rationality indicates that they make decisions with limited knowledge and resources. Given the bounded rationality and independent nature of players, a non-cooperative game is the best choice for our study of the cyber defence model. Moreover, it is a non-zero sum game because the benefit of one player is not necessarily equal to the loss of the other, i.e. the payoffs of an attacker and co-playing defender do not always add up to zero. Defenders' loss has two aspects: one is monetary and the other is non-monetary. Monetary loss includes assets value, legal fees and cost of data recovery. And, losses like reputation loss, identity theft, downtime, and business disruption are not monetary and cannot be directly equated to the attacker's benefit. These different payoffs and different strategies of the players make the game asymmetric.
\subsection{Game description and players payoffs}
We elaborate in this section the model, including the game players, their strategies and payoffs. It is a two-population  model, where players interact in a pairwise fashion. Each player has two possible strategies.
\begin{table}[H]
    \centering
    \renewcommand{\arraystretch}{1.5} 
        \caption{Summary of parameters in the model.}
    \label{table:1}
    \begin{tabular}{c|m{7cm}|c}
    \hline
    \textbf{Variables} & \textbf{Meaning of variables} & \textbf{Constraints}\\
    \hline
    $w$  & Assets value, i.e. loss to the defender for an attack & $0 < w \leq1$ \\
     \hline
     $c_a$  & Cost to the attacker for attack attempt & $ 0< c_a < w $ \\
    \hline
    $c_d$  & Cost to the defender for implementing defence & $ 0 < c_d <w $ \\
    \hline
     $b_a$  & Attacker's benefit for a successful attack & $c_a < b_a$ \\
     \hline
    $b_d$  & Defender’s benefit for not being breached & $c_d < b_d \leq w$ \\
     \hline
     $v$  & Probability of successful defence & $0 < v\leq1$ \\
     \hline
    $m$  & Probability of catching an attacker on an unsecured system & $0\leq m\leq1$ \\
     \hline
     $n$  & Probability of catching an attacker on a secured system & $0\leq n\leq1$ \\
     \hline
     $p$  & Attacker's penalty for a successful attack & $p\geq0$ \\
     \hline
     $s$  & Attacker's penalty for an unsuccessful attack & $s\geq0$ \\
     \hline
    \end{tabular}
\end{table}
Table \ref{table:1} contains the parameters and their constraints. We consider infinite, well-mixed populations for this study. The frequency of attack is denoted by $\alpha$, indicating that the attack probability of a randomly selected player from the attack population.   Similarly, $\beta$ denotes  the frequency of defence. The payoff matrix can be written as follows, in Table \ref{table:2}:
\begin{table}[H]
   \centering
   \renewcommand{\arraystretch}{1.5} 
      \caption{Payoff Matrix of the attacker and defender}
    \begin{tabular}{c|c|m{4.5cm}}
    \hline
      \textbf{Strategies} & \textbf{No Attack $(1-\alpha)$} & \textbf{Attack $(\alpha)$}\\
     \hline
     \textbf{No Defence $(1-\beta)$} & $(0,0)$ & $-w,-c_a+b_a-mp$ \\
     \hline
    \textbf{Defence $(\beta)$} & $(-c_d+b_d,0)$ & $-c_d+vb_d-w(1-v),-c_a+b_a(1-v)-vns-(1-v)mp$ \\
    \hline
  \end{tabular}
    \label{table:2}
\end{table}   
\subsection{Equilibrium and stability analysis}
After formulating the cyber-defence game, its players, strategies, and payoffs, we explore the system dynamics to find the equilibrium points and determine their stability, i.e. the state within a population where the strategy distribution remains stable overtime. This state can be calculated by using the replicator dynamics (RD) method of EGT, to find the strategy frequencies in infinite populations. 
\subsubsection{Replicator dynamics}
Replicator dynamics \cite{schuster1983replicator} is a well-established model in EGT for mathematical analysis of the idea that in any population, better-performing behavioural strategies spread \cite{roca2009evolutionary}. It is used to describe the population dynamics driven by players' payoffs to find equilibrium points of an infinite population. To study the replicator dynamics of our two-population model, we derive a set of differential equations that describe the rates of change of strategy frequencies  over time. The following set of equations shows the calculation of replicator dynamics for our model. The replicator equation for both populations at time $t$ is:
\begin{equation}
    F(x_i) = dx_i/dt = x_i(f_i - \Bar{f})
\end{equation}
where $x_i$ is the proportion of strategy $i$ in the population. $f_i$ is the fitness of type $i$ and $\Bar{f}$ is the average population fitness. The corresponding replicator equations for defender ($\beta$) and attacker ($\alpha$) populations at time $t$ are given by:
\begin{equation}
    \begin{split}
        F(\beta) = d\beta/dt = \beta(f_\beta - \Bar{f_\beta})\\
        F(\alpha) = d\alpha/dt = \alpha(f_\alpha - \Bar{f_\alpha})
    \end{split}
\end{equation}
And average fitness of the populations is as follows:
\begin{equation}
    \begin{split}
        \Bar{f_\beta} = \beta*f_\beta + (1 - \beta)f_{1-\beta}\\
        \Bar{f_\alpha} = \alpha*f_\alpha + (1 - \alpha)f_{1-\alpha}
    \end{split}
\end{equation}
The expected payoffs of no defence, $f_{1-\beta}$, and  defence, $f_\beta$, strategies in our model, can be obtained as follows:
\begin{equation}
    \begin{split}
        f_{1-\beta} &= 0 * (1 - \alpha) + (-w) * \alpha = -w\alpha\\
        f_\beta &= \alpha(-b_d+vb_d-w+vw)-c_d+b_d\\
    \end{split}
\end{equation}
Similarly, the expected payoffs of no-attack, $f_{1-\alpha}$, and attack, $f_\alpha$ strategies, are obtained below:\\
\begin{equation}
    \begin{split}
    f_{1-\alpha} &= 0 * (1 - \beta) + (0) * \beta = 0\\
    f_\alpha &= \beta(vmp-b_av-vns)-c_a+b_a–mp    
    \end{split}
\end{equation}
First, by inserting Equations (4) in eq. (2), we  obtain the replicator dynamics for the defender population.\\
\begin{equation}
    \begin{split}
       F(\beta) &= d\beta/dt = \beta(f_\beta - f'_\beta)=\beta(1-\beta)(f_{\beta}-f_{1-\beta})\\
       F(\beta) &= \beta(1-\beta)(b_d-c_d-b_d\alpha+vb_d\alpha+vw\alpha)
    \end{split}
\end{equation}  
Similarly, for the attacker population, by inserting Equations (4) in eq. (2), the replicator dynamics reads\\
\begin{equation}
    \begin{split}
     F(\alpha) &= d\alpha/dt = \alpha(f_\alpha - f'_\alpha)=\alpha(1-\alpha)(f_{\alpha}-f_{1-\alpha})\\
     F(\alpha) &=\alpha(1-\alpha)(-c_a+b_a-mp-b_a v \beta-v n s \beta+v m p \beta)
    \end{split}
\end{equation}
Now we can find equilibrium points by solving the system of two differential  equations: $ F(\beta)=0$ and $F(\alpha) = 0$. We denote $E(\beta, \alpha)$ an equilibrium point if $(\beta,  \alpha)$ is a solution. 
Clearly,  $E_1(0,0)^T$, $E_2(0,1)^T$, $E_3(1,0)^T$, $E_4(1,1)^T$, are four equilibrium points. We also have a potential internal stable point:
\begin{center}
    $E_5\left(\frac{b_a-c_a-mp}{v(b_a-mp+ns)},\frac{c_d-b_d}{vb_d-b_d+vw}\right)^T$  if  $0 < \frac{b_a-c_a-mp}{v(b_a-mp+ns)},\frac{c_d-b_d}{vb_d-b_d+vw} < 1$.
\end{center}
 
\subsubsection{Jacobian matrix}
Stability of an equilibrium point can be obtained by analysing the Jacobian matrix \cite{yang2017basic}, \cite{gandolfo1997economic} and its eigen values \cite{golub2000eigenvalue}. 
Namely, an equilibrium point is a stable equilibrium if and only if all the eigenvalues of the Jacobian matrix at that point have have negative real parts \cite{hofbauer1998evolutionary}. Jacobian matrix is formed by calculating the first-order partial derivatives of a function \cite{taras1995solution}. So, the Jacobian matrix of our model is formed as:

\[ 
\left( \begin{array}{cc}
J_{11} & J_{12}\\
        J_{21} & J_{22}\\
\end{array} \right) = \left( \begin{array}{cc}
\frac{\partial F(\beta)}{\partial \beta} & \frac{\partial F(\beta)}{\partial \alpha}\\
        \frac{\partial F(\alpha)}{\partial \beta} & \frac{\partial F(\alpha)}{\partial \alpha}\\
\end{array} \right)
\]
where
\begin{align*}
J_{11} &= \frac{\partial F(\beta)}{\partial \beta} = (1 - 2\beta) (b_d - c_d-b_d\alpha + vb_d\alpha + \alpha v w),\\
J_{12} &= \frac{\partial F(\beta)}{\partial \alpha} =  \beta(1-\beta){(b_d(v-1)+vw)},\\
J_{21} &= \frac{\partial F(\alpha)}{\partial \beta} =  \alpha(1-\alpha)(v(mp-b_a-ns)),\\
J_{22} &= \frac{\partial F(\alpha)}{\partial \alpha} = (1 - 2\alpha)(-c_a+b_a-mp-b_av\beta-vns\beta+vmp\beta).
\end{align*}
Table \ref{table:3} below shows eigen values at different equilibrium points  $E(\beta,\alpha)$. 
\begin{table}[H]
    \centering
    \renewcommand{\arraystretch}{1.5} 
    \caption{Eigen values of the Jacobian matrix}
\begin{tabular}{c|m{3.5cm}|m{3.5cm}}
    \hline
    \multicolumn{1}{c|}{\textbf{Equilibrium points}} & \multicolumn{1}{c|}{\textbf{$\lambda_1$}} & \multicolumn{1}{c}{\textbf{$\lambda_2$}}\\
    \hline
    \textbf{$E_1 (0,0)$} & $b_d-c_d$ & $b_a-c_a-mp$ \\
    \hline
    \textbf{$E_2 (0,1)$} & $c_a-b_a+mp$ & $b_dv-c_d+wv$ \\
    \hline
    \textbf{$E_3 (1,0)$} & $c_d-b_d$ & $b_a-c_a-mp-b_av-vns+vmp$ \\
    \hline
    \textbf{$E_4 (1,1)$} & $c_a-b_a+mp+b_av+vns-vmp$ & $c_d-b_dv-wv$ \\
    \hline
\end{tabular}
    \label{table:3}
\end{table}
From the observed parameter conditions and eigenvalues in Table \ref{table:3}, we can derive the initial parameter values to visualize the stability of the system at the equilibrium points. We start by assuming that the probabilities of catching and penalising the attacker are negligible as it is often agreed that the chance of catching attackers in real-world cybersecurity incidents is low \cite{wall1998catching}. 

That is, for simplicity we first set $m=0, n=0, p=0$, and $s=0$. That would simplify the formulas for $\lambda_1$ and $\lambda_2$ in Table \ref{table:3}, and we thus obtain the following general stability conditions.\\
\textbf{Stability condition 1:} The equilibrium point $E_1(0,0)$ is stable if and only if:
\begin{center}
    $b_d-c_d<0$ and $b_a-c_a<0$,
\end{center}
which simplify to
\begin{center}
    $b_d<c_d$ and $b_a<c_a$.
\end{center}
At this stable point, both the attacker and the defender are discouraged from taking any action because their costs are higher than their benefits. Namely, when these conditions hold, attackers do not attack and defenders do not invest in cybersecurity.\\

\textbf{Stability condition 2:} The equilibrium point $E_2(0,1)$ is stable if and only if:
\begin{center}
    $c_a-b_a<0$ and $b_dv-c_d+wv<0$,
\end{center}
which simplify to
\begin{center}
    $c_a<b_a$ and $c_d>v(b_d+w)$.
\end{center}
It implies that a low cost of attack motivates the attacker to attack. However, defenders are discouraged because the cost is higher than their benefit. When these conditions hold, attackers dominate the system, and defenders do not invest in cybersecurity.\\ 
\textbf{Stability condition 3:} The equilibrium point $E_3(1,0)$ is stable if and only if
\begin{center}
    $c_d-b_d<0$ and $b_a-c_a+b_av<0$
\end{center}
which simplifies to
\begin{center}
    $c_d<b_d$ and $c_a>b_a(1+v)$
\end{center}
In this condition, a low cost of defence motivates the defender to invest in security measures and protect the system. However, a high cost of attack discourages the attacker. Here, another important factor is intensity of defence $v$, increasing which will increase the expected  cost.   It isa  desirable situation in real-world when defence is so strong that either attack attempts are unsuccessful or attackers do not attack.\\  

\textbf{Stability condition 4:} The equilibrium point $E_4(1,1)$ is stable if and only if
\begin{center}
    $c_a-b_a+b_av<0$ and $c_d-b_dv-wv < 0$
\end{center}
which simplify to 
\begin{center}
    $c_a<b_a(1-v)$ and $c_d<v(b_d+w)$
\end{center}
Low costs and high benefits encourage the players to act. So, the attacker will continue to attack and the defender will always be defending. If both conditions hold, both attackers and defenders remain active, leading to highly contested environment where defence and attack are persistent.\\

\textbf{Internal equilibrium:} The model has one potential  internal stable equilibrium point at $E_5\left(\frac{b_a-c_a-mp}{v(b_a-mp+ns)},\frac{c_d-b_d}{vb_d-b_d+vw}\right)$. However, for all the game configurations considered in this work, whenever $0 < \frac{b_a-c_a-mp}{v(b_a-mp+ns)},\frac{c_d-b_d}{vb_d-b_d+vw} < 1$ is satisfied, the two eigen values do not have a negative real part at the same time. 

Due to the complexity of the explicit forms of the eigen values at the internal equilibrium, we do not provide analytical conditions for its stability, and use numerical simulations for its calculation instead.
 
All eigen values in Table \ref{table:3} depend on cost and benefit to the players, and a change in these parameters will affect players' behaviour. Real-world scenarios support the above equilibrium solutions where certain comparative relationships among variables impact the system stability. Cyber systems contain valuable data which must be protected and attackers are always interested in confidential information to gain benefits. Organizations who do not care about strong security measures, leave themselves exposed to cyber threats.

To illustrate the stability conditions discussed above, the phase diagrams in Figure \ref{fig:4} and Figure \ref{fig:5} are plotted. Parameter values are chosen based on the constraints given in Table \ref{table:1}. 

\begin{figure}[H]
\centering
\begin{subfigure}{0.45\textwidth}
\includegraphics[width=1\linewidth, height=5cm]{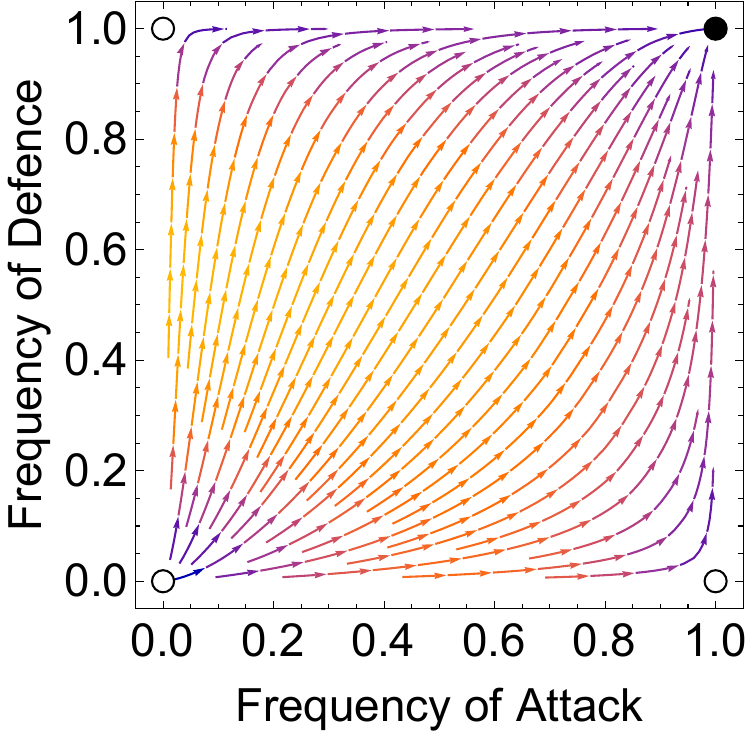}
\caption{%
    \begin{tabular}{@{}ll@{}}
    $b_a=0.90$, & $b_d=0.79$ \\
    $c_a=0.51$, & $c_d=0.20$ \\
    $w=0.98$,   & $v=0.26$   \\
    \end{tabular}%
}
\end{subfigure}
\begin{subfigure}{0.45\textwidth}
\includegraphics[width=1\linewidth, height=5cm]{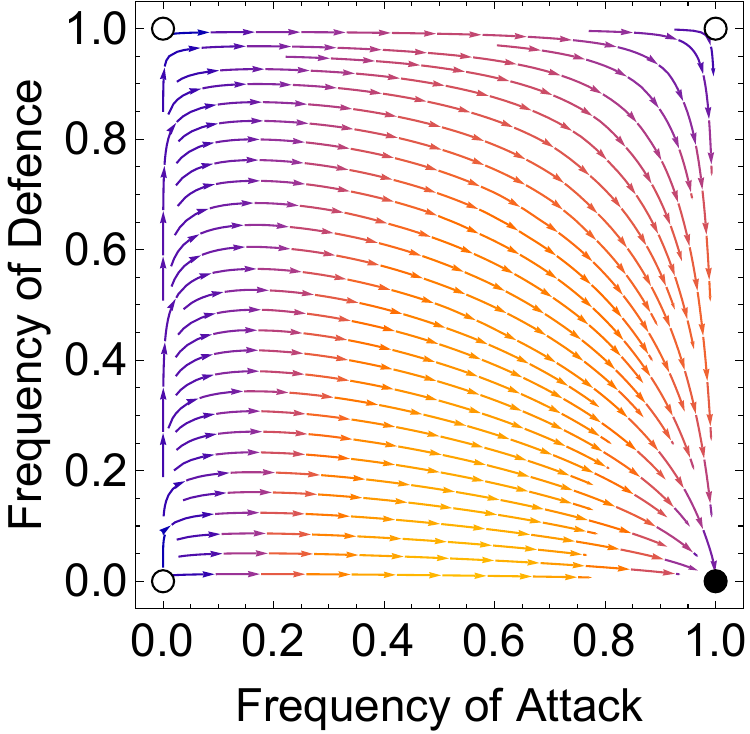}
\caption{%
    \begin{tabular}{@{}ll@{}}
    $b_a=0.52$, & $b_d=0.37$ \\
    $c_a=0.29$, & $c_d=0.34$ \\
    $w=0.43$,   & $v=0.24$   \\
    \end{tabular}%
}
\end{subfigure}
\caption{Phase plot for stability of $E_4=(1,1)$ and $E_4=(1,0)$ in subplots (a) and (b) respectively. Solid circles denote the stable point and blank circles are denoting unstable points in the system.}
\label{fig:4}
\end{figure}
Phase plots in Figure \ref{fig:4} show the stability of equilibrium points when defender is defending and attacker may or may not attack, because we are concerned about the effective defence and its stability in cyber world. 
\begin{figure}[H]
\centering
\begin{subfigure}{0.45\textwidth}
\includegraphics[width=1\linewidth, height=5cm]{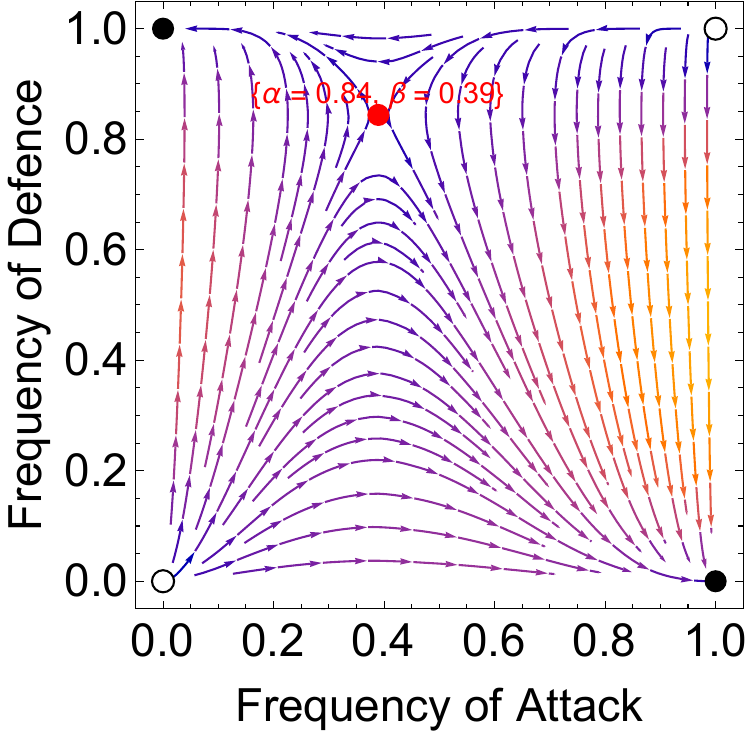}
\caption{%
    \begin{tabular}{@{}ll@{}}
    $b_a=0.79$, & $b_d=0.72$ \\
    $c_a=0.69$, & $c_d=0.54$ \\
    $w=0.98$,   & $v=0.15$   \\
    \end{tabular}%
}
\end{subfigure}
\begin{subfigure}{0.45\textwidth}
\includegraphics[width=1\linewidth, height=5cm]{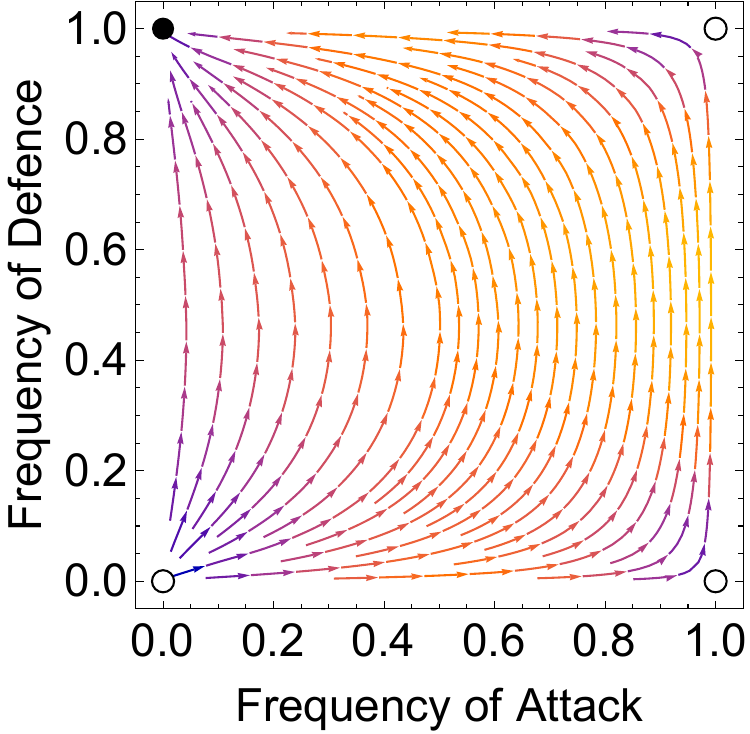}
\caption{%
    \begin{tabular}{@{}ll@{}}
    $b_a=0.24$, & $b_d=0.47$ \\
    $c_a=0.18$, & $c_d=0.41$ \\
    $w=0.47$,   & $v=0.54$   \\
    \end{tabular}%
}
\end{subfigure}
\caption{Phase plot for stability of internal  equilibrium and $E_2=(0,1)$ in subplots (a) and (b) respectively. Solid circles denote the stable point and blank circles are denoting unstable points in the system.}
\label{fig:5}
\end{figure}
An example plot for internal equilibrium points is shown in Figure \ref{fig:5} subplot (a), with eigen values ${\lambda_1=0.041501,\lambda_2=-0.041501}$. Subplot (b) shows no defence and attack stability.
\section{System dynamics and stability analysis}
After mathematically deriving equilibrium points, now we use them to explore the system dynamics. To systematically  understand the cyber attack and defence model and its evolutionary stability, we randomly sampled   100,000  games from the parameters space, which satisfy the constraints given in Table \ref{table:1}. This random game approach has been shown useful to examine the overall complexity and dynamics of the system behaviour and when the game payoff matrix is not deterministic due to environment changes and noise   \cite{duong2025evolutionary,han2012equilibrium,rand2011evolution,han2016emergence,galla2013complex}. For each game, we calculate  equilibrium points and then determine their stability. 
This generated random game data is used to plot the ratio of equilibrium points, the impact of defence intensity on the strategies, the impact of parameters on stability, and their relationships.  
All these aspects are discussed subsequently.
\subsection{Number of stable equilibrium points}
In the generated random game configurations we get different stability behaviours. As there are several possible equilibrium points, we examine if a game might have  more than one stable point simultaneously, as the number of stable points indicates can predict a co-existence of different types in a population and the maintenance of polymorphism \cite{han2012equilibrium,duong2016analysis,broom1997multi}. Figure \ref{fig:6} subplot(a) shows, 98.4\% games have a single equilibrium point and only a few configurations show complex dynamics i.e., 1.6\% games having two equilibrium points. It reflects the tendency of the system to converge toward a single stable state under random conditions. It is worth noting that more than two equilibrium points are not stable at the same time.
Here are the images to show the behaviour. 
\begin{figure}[H]
\centering
\begin{subfigure}{0.48\textwidth}
\includegraphics[width=1\linewidth, height=4cm]{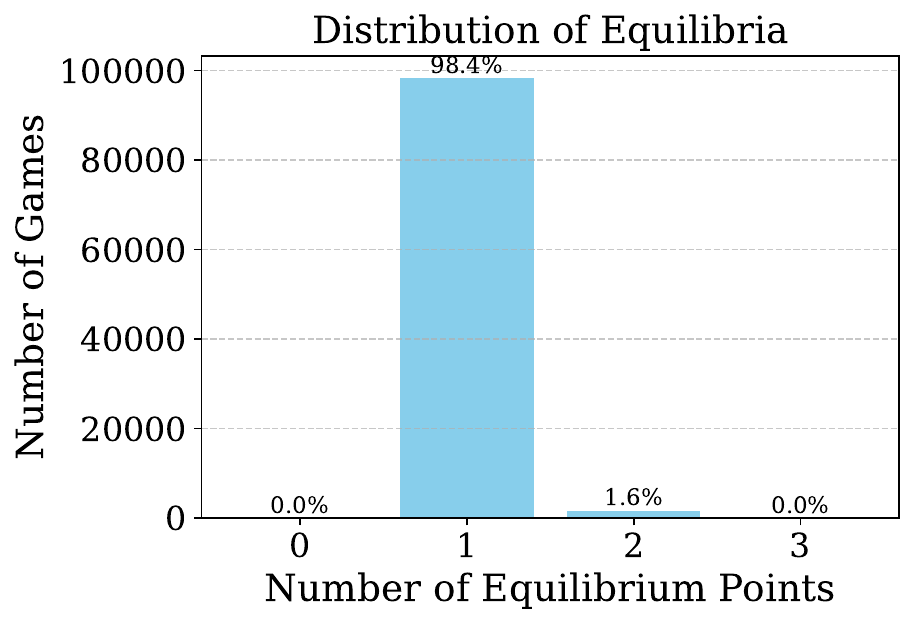} 
\caption{}
\end{subfigure}
\begin{subfigure}{0.48\textwidth}
\includegraphics[width=1\linewidth, height=4cm]{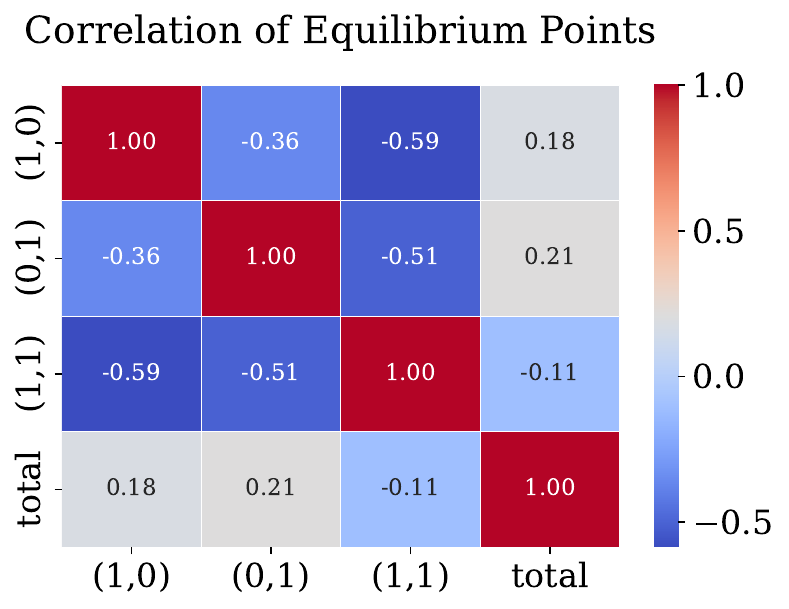}
\caption{}
\end{subfigure}
\caption{Subplot (a) shows the distribution of equilibrium points in randomly generated game configurations. Subplot (b) shows correlation matrix between the equilibrium points}
\label{fig:6}
\end{figure}
The heatmap representation of the correlation matrix in subplot (b) shows that a strong negative correlation exists between $(1,1)$ and other equilibrium points which is the reason three equilibrium points are never stable simultaneously. $(1,0)$ and $(0,1)$ have -0.36 correlation, suggesting a weaker inverse relationship. We can see in a few configurations both points are stable simultaneously. Moreover, these points $(1,0)$ and $(0,1)$ have a positive correlation 0.18 and 0.21 respectively, with total stability, showing their slight contribution to the total value. Figure \ref{fig:5} shows an example of a game configuration with two stable equilibria  at $(1,0)$ and $(0,1)$. Depending on the initial condition regarding the frequency of attack and defence, one of the stable points will be reached. 
\subsection{Frequency of stable equilibria in random games}
It is useful to understand what equilibrium point is more likely to be stable, when the game payoff matrix is randomly drawn. This is particularly useful when a prediction needs to be made under uncertainty, which can happen when the environment changes rapidly, or when human errors are frequent  \cite{duong2025evolutionary}. 
\begin{figure}[H]
    \centering
    \includegraphics[scale=0.35]{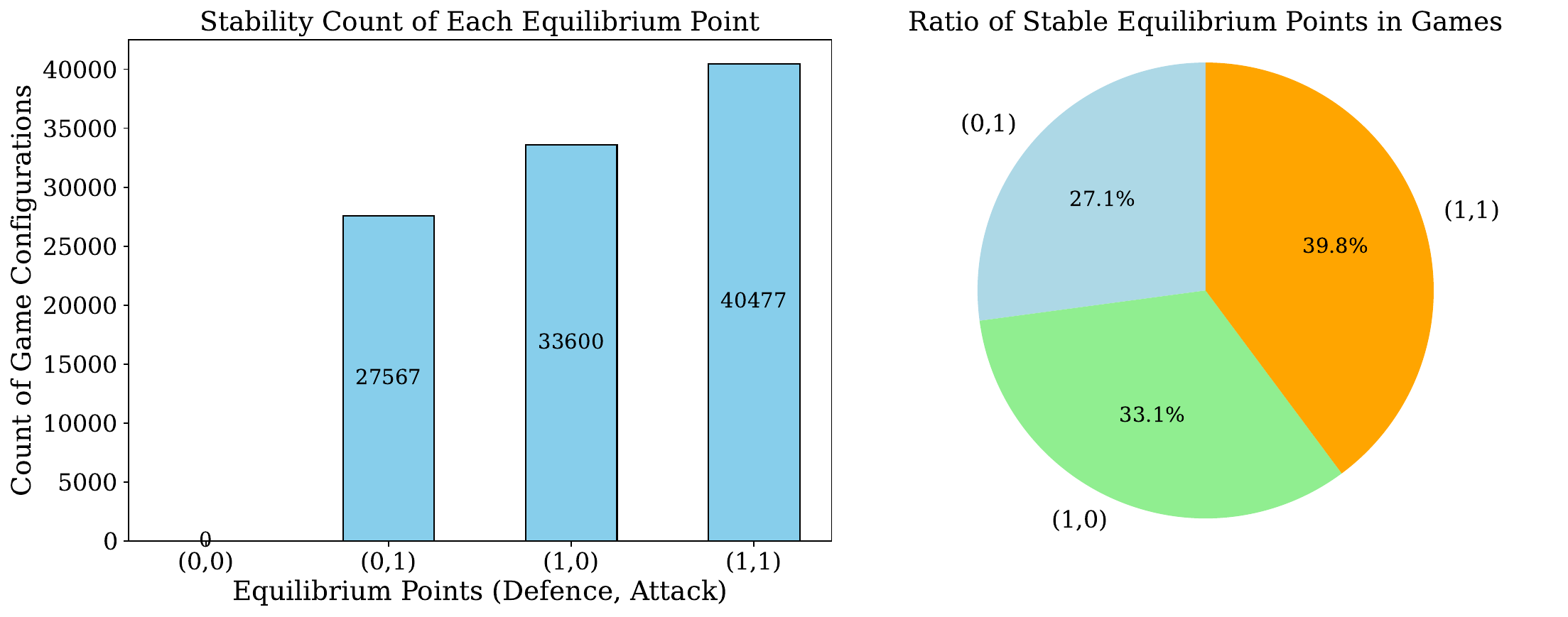}
    \caption{Subplot (a) shows the count of game configurations with each equilibrium point. $E_4(1,1)$ shows dominance with highest number of occurrences, followed by $E_3(1,0)$ and $E_2(0,1)$. The pie chart in subplot (b) shows the ratio of equilibrium points.} 
    \label{fig:7}
\end{figure}
Figure \ref{fig:7}$(a)$ shows  the number of game configurations on y-axis with at least one stable point and the equilibrium points on x-axis respectively. And Figure \ref{fig:7}$(b)$  shows the ratios of game configurations for each stable equilibrium point. The equilibrium point $E_4(1,1)$ shows the highest ratio of stability in $39.8\%$ game configurations, showing a continuous war between defender and attacker and indicating the need for a strong defence strategy. The next high ratio is for the equilibrium point $E_3(1,0)$, when defender defends and no attack happens. The existence of equilibrium point $E_2(0,1)$ demonstrates that unprotected systems are also attacked. The equilibrium point $E_1(0,0)$ does not exist, indicating that in cyberspace, running an unsecured system is not recommended. These subplots provide a clear understanding of how the majority of stable systems tend to favour defence strategies to achieve stability.

\subsection{Defence probability and strategy count}
From the above analysis of stability ratio, we observed that always defend and attack is the dominant collective behaviour in cyber-defence system. It highlights the need for strong defensive measures (i.e. high $(v)$) to protect the system from adversaries. In Figure \ref{fig:8}, we analyse the number of game configurations showing stability for each equilibrium point, for varying the defence probability $v$.
\begin{figure}[H]
\centering
\begin{subfigure}{0.32\textwidth}
\includegraphics[width=1\linewidth, height=4cm]{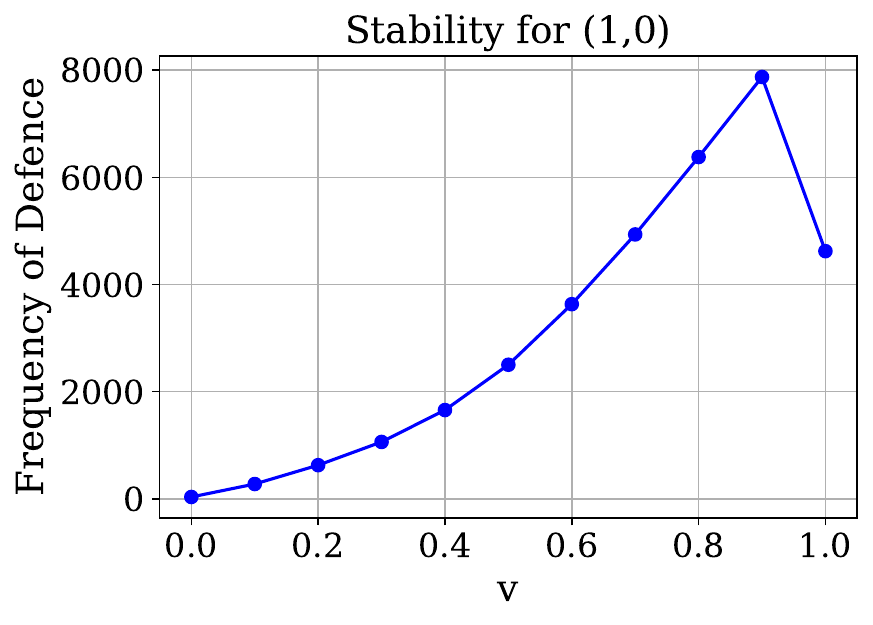} 
\caption{}
\end{subfigure}
\begin{subfigure}{0.32\textwidth}
\includegraphics[width=1\linewidth, height=4cm]{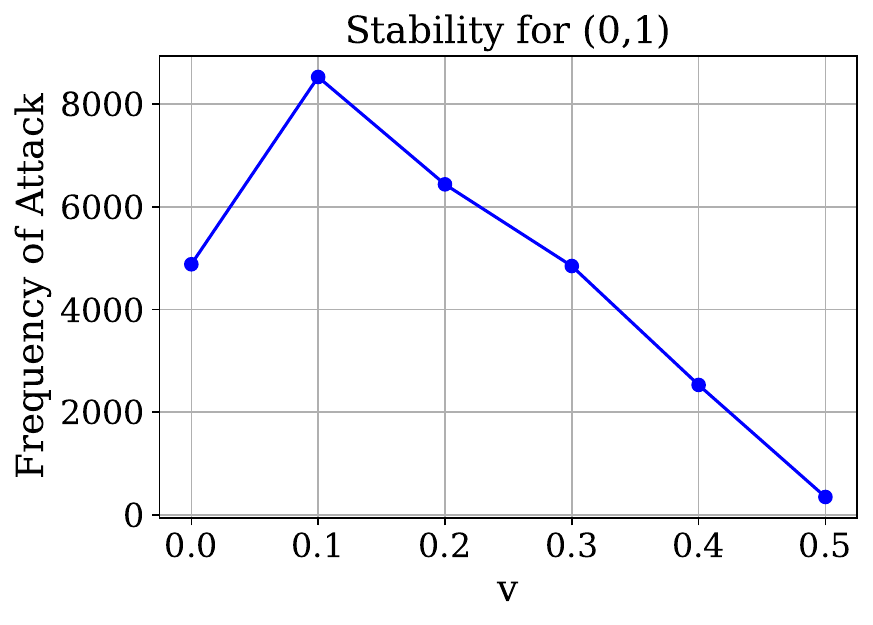} 
\caption{}
\end{subfigure}
\begin{subfigure}{0.32\textwidth}
\includegraphics[width=1\linewidth, height=4cm]{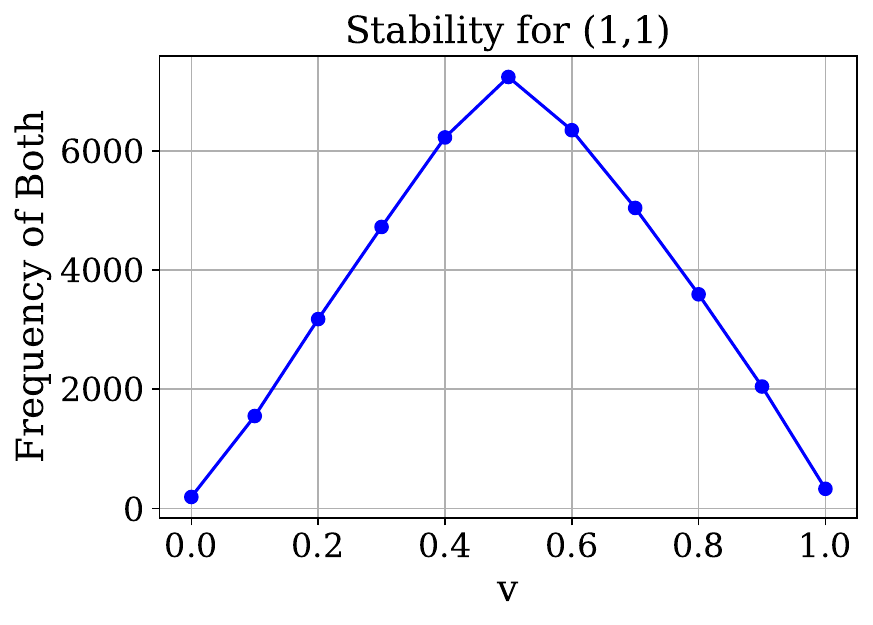}
\caption{}
\end{subfigure}
\caption{Frequencies of equilibrium points (1,0), (0,1), and (1,1) as a function of the defence intensity  $v$,   in the set of randomly sampled games.}
\label{fig:8}
\end{figure}
Figure \ref{fig:8}$(a)$,  shows  that as the probability of defence $v$ increases the frequency of $(1,0)$ (pure defender strategy) increases. So, stability is associated with strong defence implementations.
It is worth noting that $v$ has a peak around  0.9, i.e. 90\% probability of defence, suggesting that perfect defence does not guarantee complete stability and it may make the system unstable. The reason behind it is that 100\% intensity of defence $v$ is costly to achieve as a very high defence cost is unfavourable for the defenders. So, defender needs to ensure effective defence without overwhelming the resources.
The subplot $(b)$ shows the frequency of $(0,1)$ (pure attacker strategy) decreases as $v$ increases i.e., starts at a high frequency for low $v$, peaking at $v=0.1$, and then declines rapidly. For, $v>0.5$, attacker rarely dominates, as v increases, successful defence becomes more effective, making pure attack strategies unstable. 
Subplot $(c)$ indicates that frequency of $(1,1)$ (defence and attack) follows a bell shaped curve peaking at $v=0.5$. So, at low $v$, defence is weak, pure attack strategies dominate. At mid range values of v, both strategies coexist, maximising $(1,1)$ stability. After $v=0.5$, as defence is stronger, attacks are discouraged, shifting the system towards pure defence dominance. This reinforces the idea that more secure systems are less prone to attacks.

\subsection{Impact of parameters on system stability}
The cyber defence system and its stability depend on some parameters, and following the previous analysis, we now explore how the costs and benefits to the players, assets value, and probability of defence influence the system's stability at the equilibrium point $(1,1)$. Moreover, to validate the applicability of our model, we have used a publicly available cybersecurity dataset \cite{tsen2022exploratory} about several attacks reported from 2014-2020, and we found fascinating insights. This validation is important because it connects the theoretical results to real-world scenarios, enhancing the applicability of this knowledge-based approach.  
\begin{figure}[H]
\centering
\begin{subfigure}{0.45\textwidth}
\includegraphics[width=1\linewidth, height=4cm]{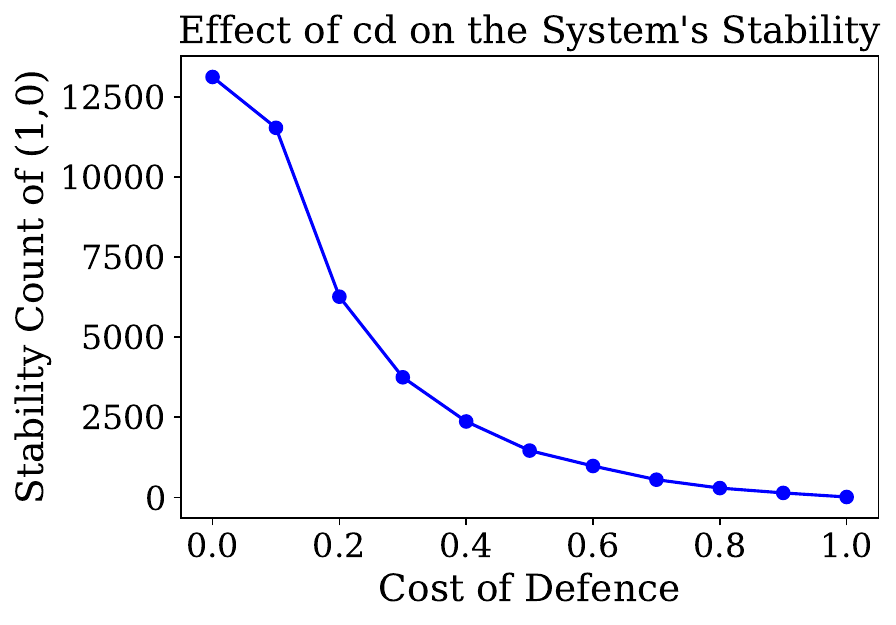} 
\caption{}
\end{subfigure}
\begin{subfigure}{0.45\textwidth}
\includegraphics[width=1\linewidth, height=4cm]{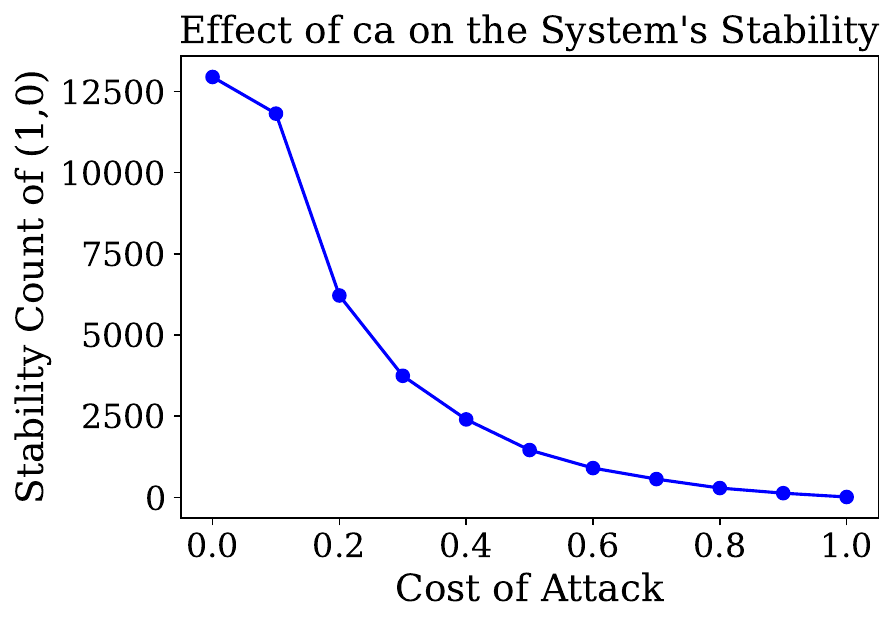} 
\caption{}
\end{subfigure}
\caption{Subplots $(a)$ and $(b)$ show the number of game configurations decrease sharply with an increase in the defender and attacker costs.}
\label{fig:9}
\end{figure}
\subsubsection{Cost of defence}
As depicted in Figure \ref{fig:9}$(a)$, when the cost of defence $c_d$ increases, the number of game configurations stable at $(1,1)$ decreases. This indicates that the higher cost of implementing defence directly reduces the organisation's ability to sustain defence. Similarly, in real-world scenarios, smaller companies can not invest more  security implementation, as compared to larger companies, and face more attack attempts as shown in Figure \ref{fig:10}. The reasons include the lack of budget and expertise needed for stronger defence implementations and the attackers exploit this vulnerability \cite{sangani2012cyber}. While larger organizations invest in layered defences and keep the highly confidential data nearly inaccessible. There is a need for a reduced cost of defence $c_d$ otherwise, it becomes harder to maintain stability, likely due to the limited resources available for effective defence.
\begin{figure}[H]
    \centering
    \includegraphics[scale=0.40]{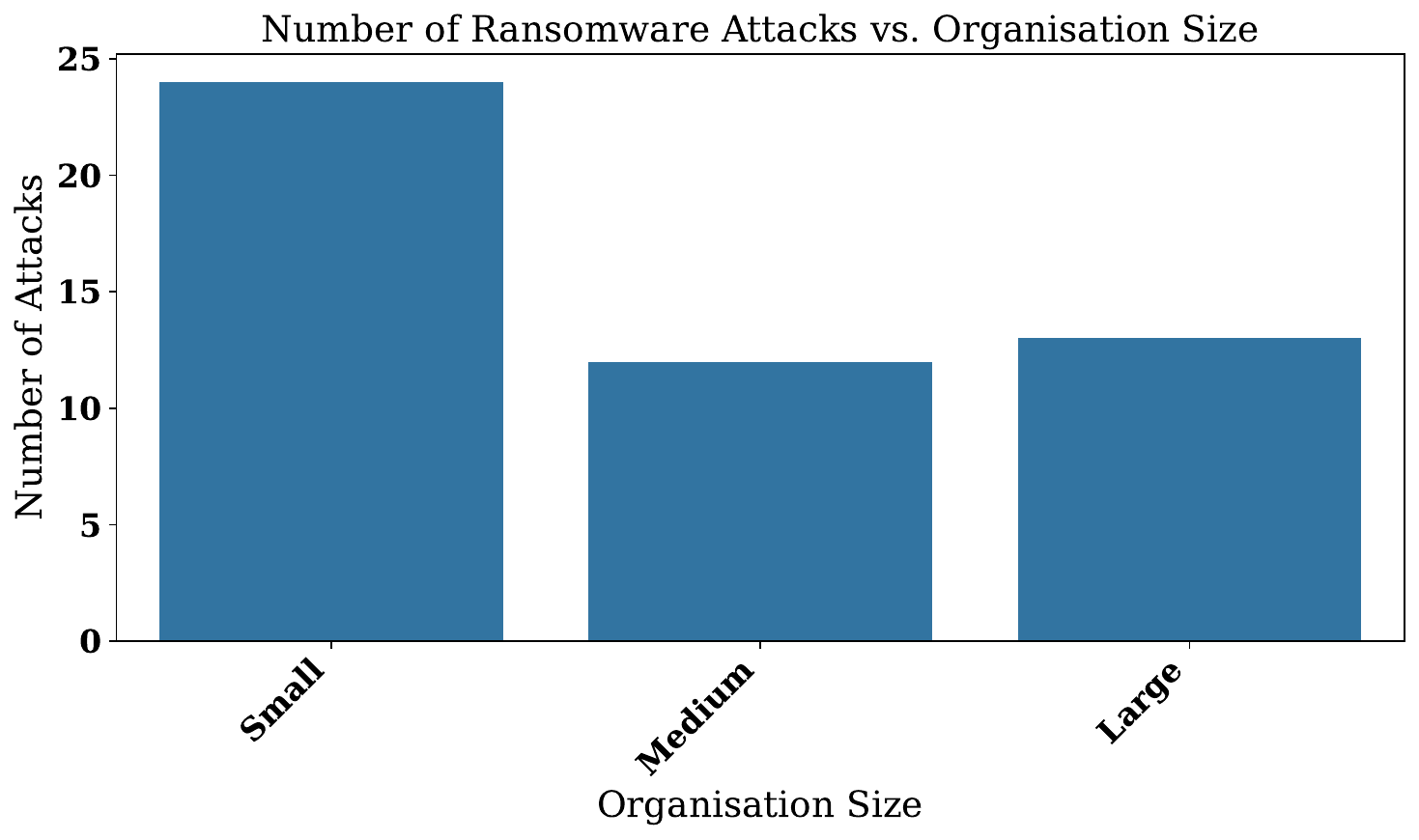}
    \caption{Small organizations are targeted by cyber-attacks more than larger ones.} 
    \label{fig:10}
\end{figure}
\subsubsection{Cost of attack}
Cost of attack $c_a$ also shows a similar trend like the cost of defence and in Figure\ref{fig:9}, $(b)$ illustrates that the number of stable game configurations is high when attack costs are low. This reflects how less costly attacks are more likely to happen, while higher attack costs discourage attackers and defence dominates.  
\begin{figure}[H]
    \centering
    \includegraphics[scale=0.40]{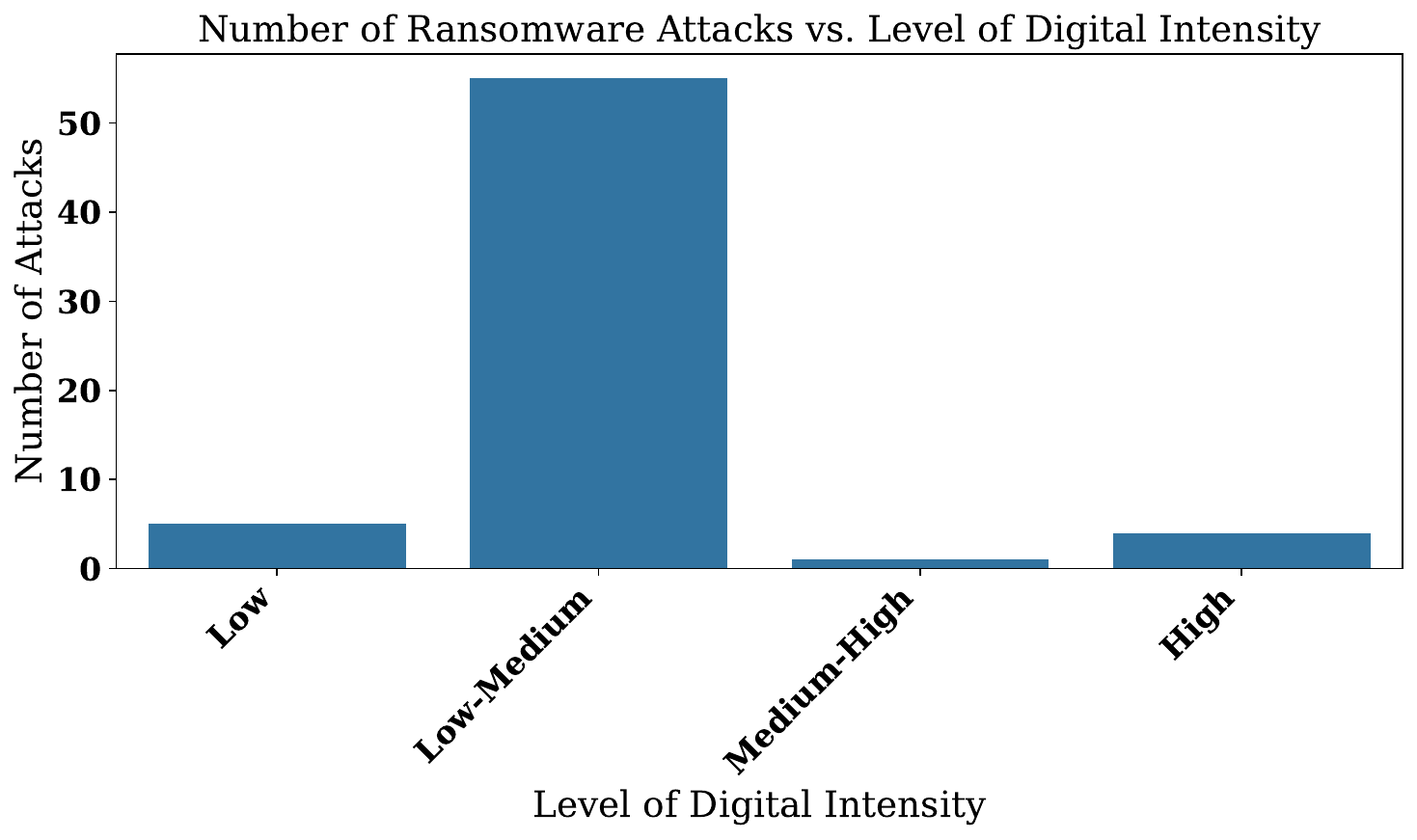}
    \caption{Organizations with low-medium digital intensity face more attacks.} 
    \label{fig:11}
\end{figure}
The graph in Figure \ref{fig:11} reinforces the idea from real-world attack data to show the attacks are prevalent in organisations with low to medium digital intensity due to their partial digitisation and lack of sufficient security measures, making attacks less costly. In contrast, high digital-intensity organisations, such as large firms with robust cybersecurity systems, experience fewer attacks as these environments are expensive to exploit, aligning with our model’s prediction. 
\begin{figure}[H]
\centering
\begin{subfigure}{0.45\textwidth}
\includegraphics[width=1\linewidth, height=4cm]{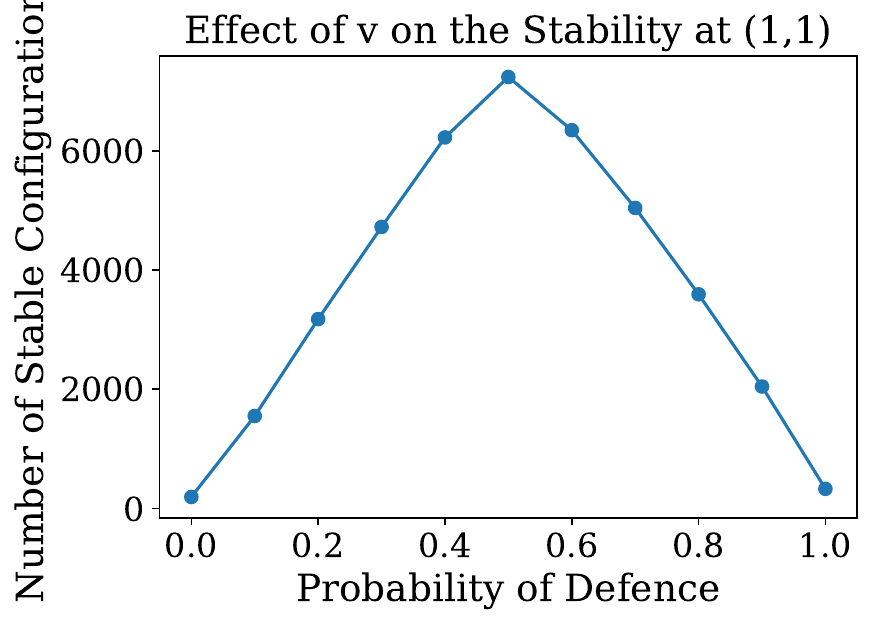} 
\caption{}
\end{subfigure}
\begin{subfigure}{0.45\textwidth}
\includegraphics[width=1\linewidth, height=4cm]{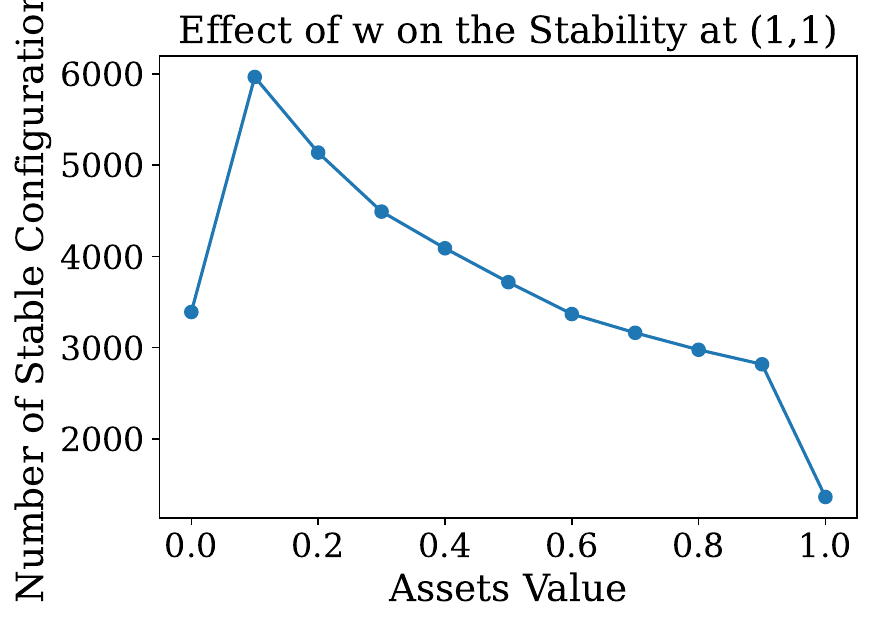} 
\caption{}
\end{subfigure}
\caption{Subplots $(a)$ and $(b)$ show the number of game configurations vs probability of defence $v$ and the asset value $w$ respectively.}
\label{fig:12}
\end{figure}
\subsubsection{Probability of successful defence \texorpdfstring{$v$}{v}}
Figure \ref{fig:12}(a) shows that the probability of defence $(v)$ has a direct impact on the number of stable game configurations. When $v$ is very low, stability is also low because of no or little defence of the system. When $v$ is very high, stability decreases, because it will become difficult to maintain high intensity of defence as it increases the cost of defence. This analysis suggests that moderate levels of $v$, according to the size of the organisations, lead to highest stability as it is manageable and is more likely to repel attacks effectively.
\begin{figure}[H]
    \centering
    \includegraphics[scale=0.40]{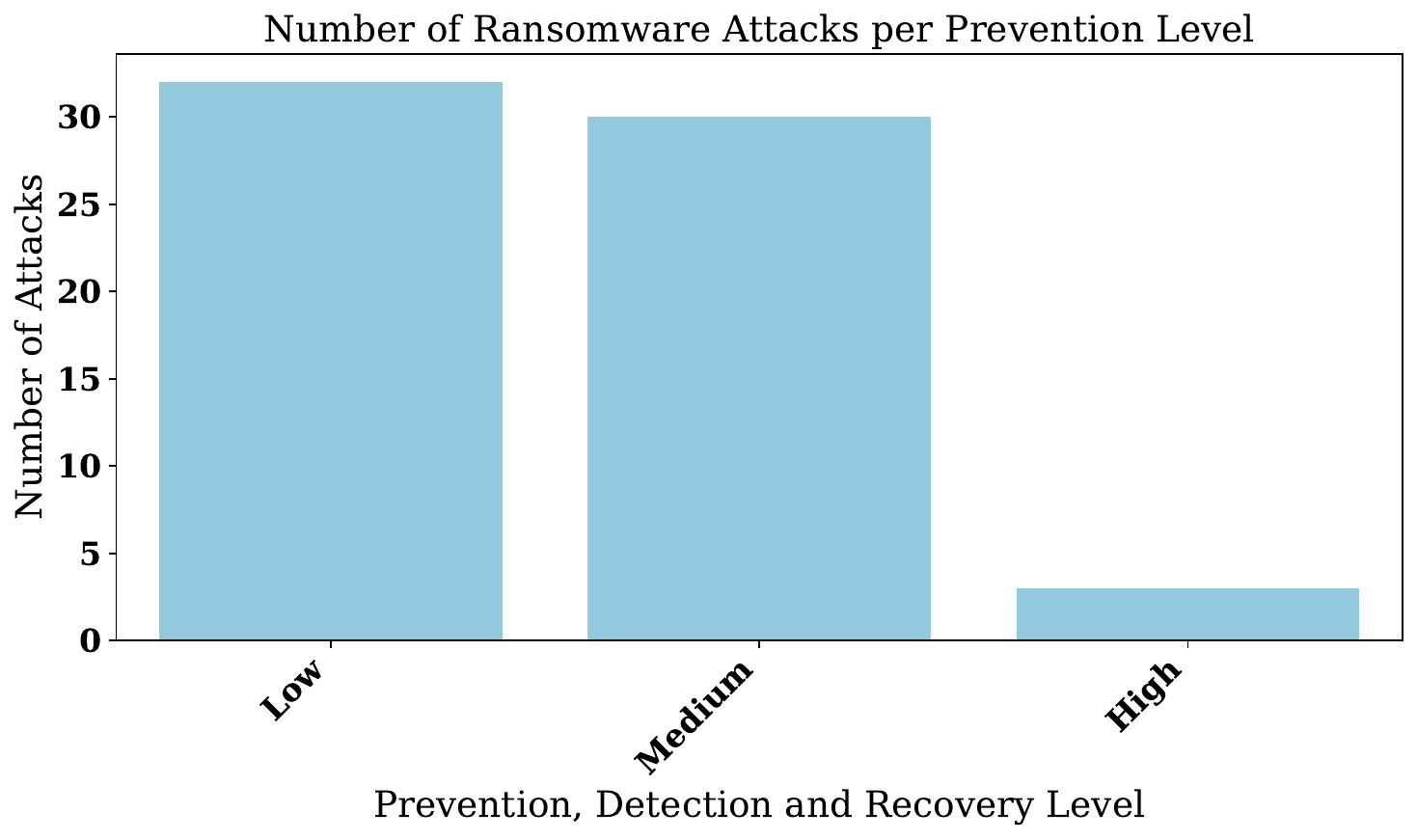}
    \caption{Organisations with high prevention, detection and recovery levels face less successful attacks.} 
    \label{fig:13}
\end{figure}
The positive correlation between the probability of defence and system stability mirrors real-world trends as in Figure \ref{fig:13}. Organisations with high prevention levels, such as those employing advanced threat detection and zero-trust architecture, experience fewer successful attacks.  For instance, large corporations like Google or Amazon, have implemented robust defences, making them less vulnerable compared to smaller enterprises with lower prevention levels.

Figure \ref{fig:12}(b) shows that for increasing assets value $w$, the number of stable configurations increase initially, but decreases as $w$ continues to rise. This is because attacks will increase for manageable costs but for a very high $w$ the number of stable configurations drops significantly.
\begin{figure}[H]
\centering
\begin{subfigure}{0.45\textwidth}
\includegraphics[width=1\linewidth, height=4cm]{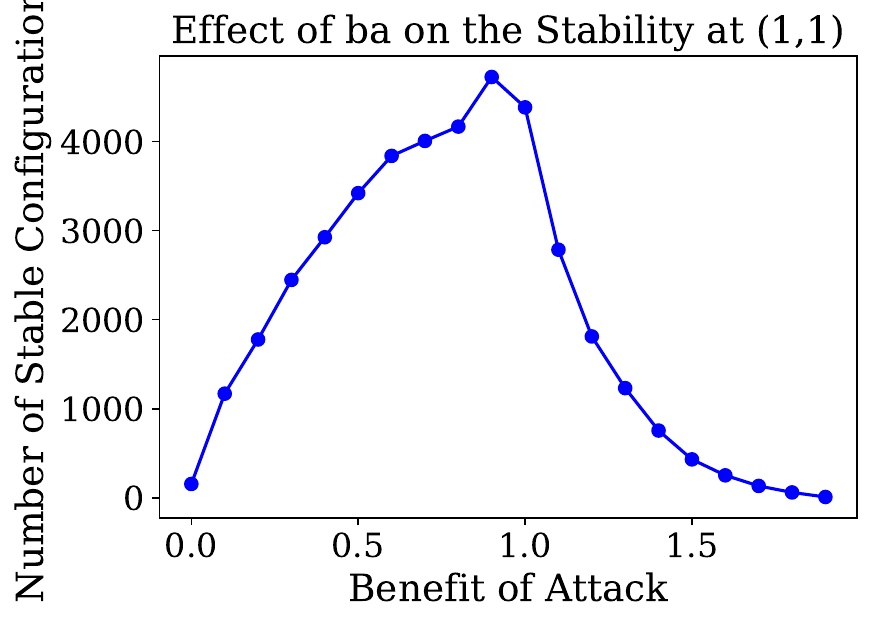} 
\caption{}
\end{subfigure}
\begin{subfigure}{0.45\textwidth}
\includegraphics[width=1\linewidth, height=4cm]{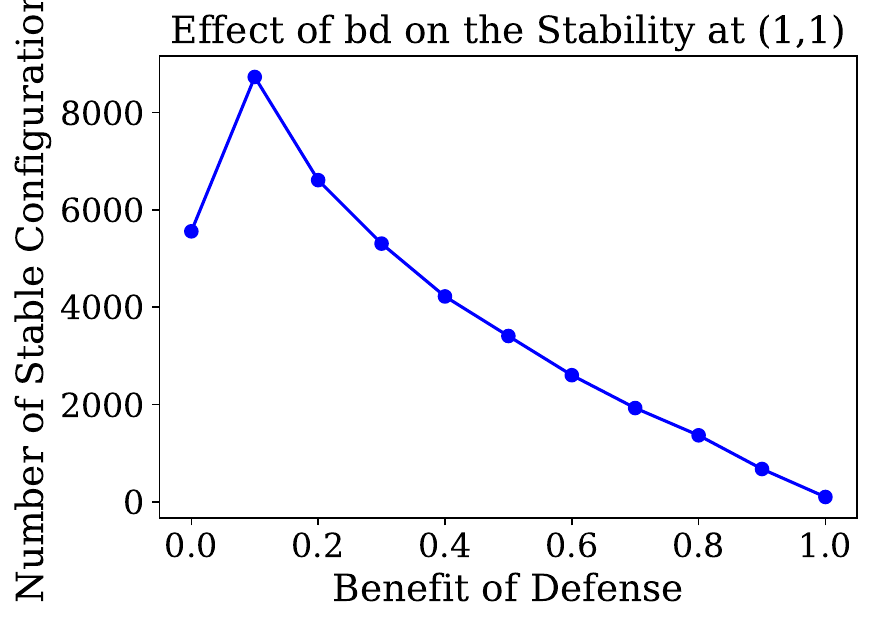} 
\caption{}
\end{subfigure}
\caption{Subplot $(a)$ shows a peak around moderate values of $b_a$. Subplot $(b)$ shows a decrease in stable game configurations with an increase in $b_d$.}
\label{fig:14}
\end{figure}
Figure \ref{fig:14}$(b)$ indicates that low benefit of attack $b_a$ makes attacking less rewarding and encourages stability at $(1,1)$. As $b_a$ increases, attack becomes beneficial to attacker and at very high $b_a$, the attacker gains significant advantage. But, when the attacker's benefit becomes higher than the asset value, stability count reduces because these cases are rare. Subplot $(b)$ shows a decrease in the stability count for an increase in benefit of defence $b_d$. This is because defending a costly online platform is also very costly as attackers always try to attack these systems. This reduces the game counts significantly for a very high value of $b_d$. This is because the cost of sustaining such a strong defence might outweigh the benefits. 
\subsection{Impact of attacker fine on equilibria}
We have kept probability of catching and penalizing the attacker equal to zero so far and analysed the cyber attack and defence scenario because in real-world, catching and penalizing cyber attacker is often not possible. But, there is a possibility of catching the cyber attackers and laws to penalise them if they are under the jurisdiction \cite{oh2014need}. We now study the impact of non-negligible probabilities of catching and penalizing the attacker in case of unsuccessful and successful attacks. In Table \ref{table:1}, we have defined variables $m$ and $p$ as the probabilities of catching the attacker on an unsecured system and its penalty for a successful attack, respectively and variables $n$ and $s$ as the probabilities of catching the attacker on a secured system and its penalty for an unsuccessful attack, respectively. For simplicity, we are denoting these four variables by using two variables $f_u$ and $f_s$ where $f_u = n*s$ and $f_s=m*p$, i.e., fine for unsuccessful and successful attack respectively. We have considered two scenarios where fines are equal to 0.1 in Figure \ref{fig:15} and 0.5 in Figure \ref{fig:16} to see the impact on system stability and we observed a drastic change in the equilibrium strategies. The system transitions from a stable state where defence and attack coexists to a new equilibrium where defenders are successfully defending the system and attackers are deterred from launching attacks. 
\begin{figure}[H]
    \centering
    \includegraphics[scale=0.32]{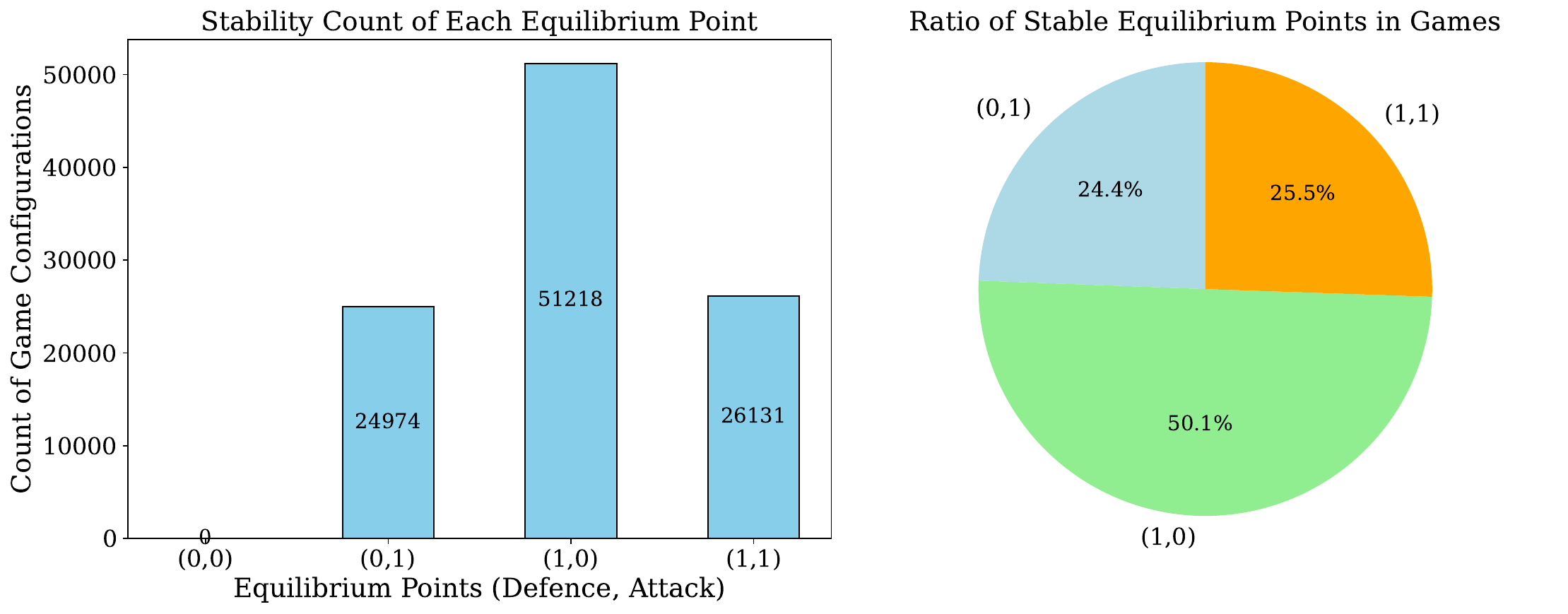}
    \caption{Subplot (a) shows the count of game configurations at each equilibrium point when fines to the attacker i.e., $f_u=0.1$ and $f_s=0.1$. $E_3(1,0)$ occurs most frequently, followed by $E_4(1,1)$ and $E_2(0,1)$. The pie chart in subplot (b) shows the ratio
of equilibrium points.} 
    \label{fig:15}
\end{figure}

\begin{figure}[H]
    \centering
    \includegraphics[scale=0.32]{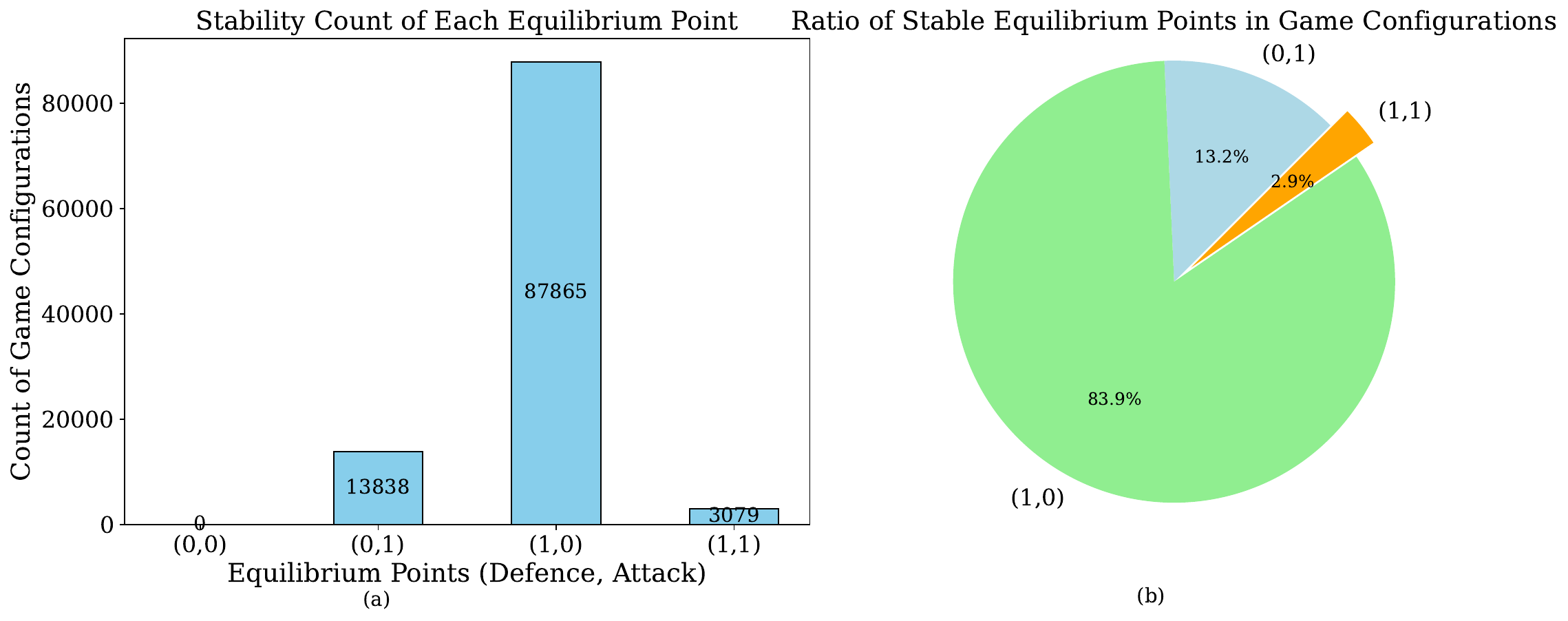}
    \caption{Subplot (a) shows the count of game configurations at each equilibrium point when fines to the attacker are $f_u=0.5$ and $f_s=0.5$. $E_3(1,0)$ occurs  most frequently, followed by $E_2(0,1)$ and $E_4(1,1)$. The pie chart in subplot (b) shows the ratio
of equilibrium points.} 
    \label{fig:16}
\end{figure}
The equilibrium shift in Figures \ref{fig:15} and \ref{fig:16} suggests that even a small chance of enforcement can have significant impact and reduce attackers' interest in attacking by reducing their incentives from malicious activities. Attackers' behaviour can be influenced by deterring them \cite{tran2018law}. Their interest in such harmful activities can be reduced by making them think that they will face punishment and a very high fine of attack. External parties such as government, institutions, law enforcement agencies, and regulatory bodies  can estimate the amount of fine according to the severity of the crime \cite{khadam2023punish} and ensure its applicability to the violators. These findings align with the existing literature on game-theoretic models, highlighting the role of strong defensive measures in shaping adversarial behaviour \cite{8422719}.

We can see that the theoretical analysis of our evolutionary game model is consistent with the real-world cyber-attack dataset, proving the significance of our study. The findings of this study highlight the importance of intelligent resource allocation \cite{sokri2018optimal}, aligning with the knowledge-based systems approach to optimize cybersecurity strategies. To stay ahead of cybercriminals, businesses must focus on establishing effective defence measures before an attack happens. Organizations should learn from these previous attack trends to predict when attackers will act in the future and take early protective measures. This means that working together, raising awareness, sharing knowledge, and developing systems will keep cyberspace safer in the long run \cite{bada2019developing}.

\section{Social Welfare}
Building upon the equilibrium analysis presented in the previous sections, it becomes essential to understand how well the cyber-defence system works as a whole. Social welfare serves as a key metric in this regard, allowing us to identify strategies that maximise the collective benefit of the population \cite{10.1145/3375627.3375829,han2024evolutionary}. This is particularly important in non-cooperative cyber defence scenarios to better understand the balance between strategic stability and societal benefit. 

The social welfare of a strategy pair \cite{apt2014selfishness} can be defined as:
\begin{equation}
   SW(s) := \sum_{i=1}^{2} u_i(s),
   \label{eq:summation}
\end{equation}
where s is the strategy chosen by player $i$ and $u_i$ is the player's payoff in that strategy pair. In our cyber-defence model, social welfare for each possible stable state $(0,1)$, $(1,0)$, and $(1,1)$, consisting of strategy pair $(\beta, \alpha)$ is computed as the sum of defender and attacker payoffs. For instance, the social welfare for the parameter values used in Figure \ref{fig:4} $(a)$ can be calculated as given in Table \ref{table:4}:
\begin{table}[H]
    \centering
    \caption{Social Welfare}
    \label{table:4}
    \begin{tabular}{c|c}
    \hline
      \textbf{Strategy pairs $(\beta,\alpha)$}& \textbf{Social Welfare}\\
     \hline
     No Defence, No Attack & 0 \\
     \hline
     No Defence, Attack & -0.59 \\
     \hline
     Defence, No Attack & 0.59 \\
     \hline
     Defence, Attack & -0.56 \\
     \hline
  \end{tabular}
\end{table}
It can be seen  that the strategy pair (Defence, No Attack) has the highest payoff and highlights the efficiency of unilateral defence in deterring attacks and maximizing societal benefit under the given parameters. 
Now, we calculate the statistical values for the social welfare across random games. Here is the comparison of statistical values of all equilibrium points to show which equilibrium point occurs the most.
Now, to see the trend of defender and attacker strategies, we can plot the social welfare for our random games. 
\begin{figure}[H]
\centering
\begin{subfigure}{0.40\textwidth}
\includegraphics[width=1\linewidth, height=4cm]{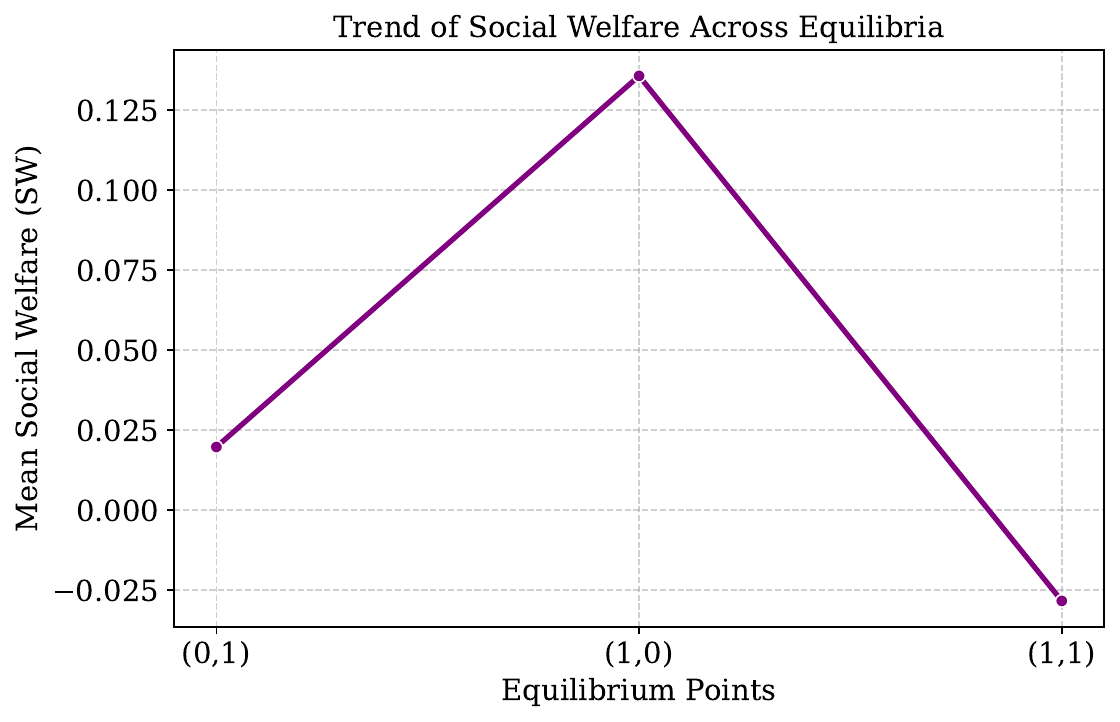}
\caption{}
\end{subfigure}
\begin{subfigure}{0.55\textwidth}
\includegraphics[width=1\linewidth, height=4cm]{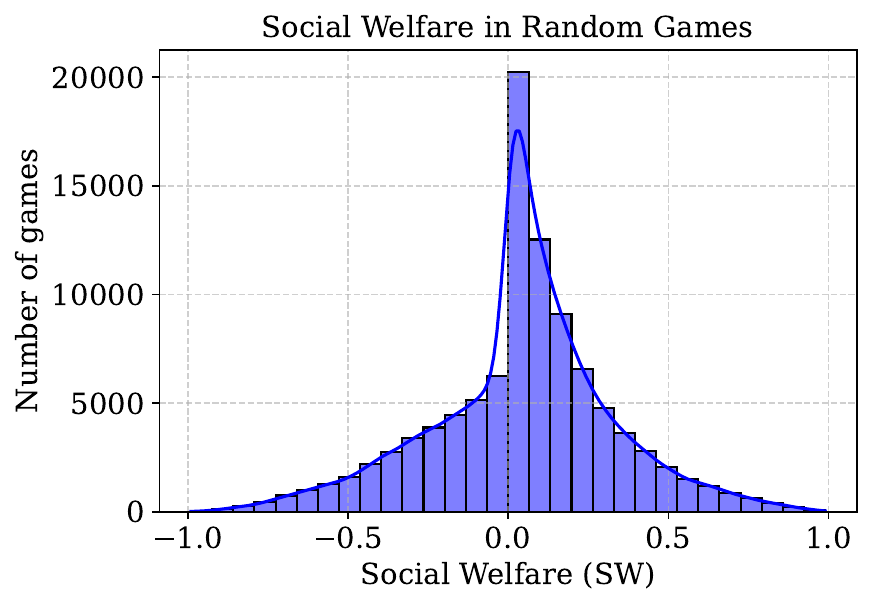}
\caption{}
\end{subfigure}
\caption{Subplots (a) shows the trend of social welfare across the equilibrium points and (b) shows the social welfare distribution in random games.}
\label{fig:17}
\end{figure}
Figure \ref{fig:17} $(a)$ shows the average value of social welfare for each strategy pair. It shows highest social welfare can be achieved when defender defends and attacker do not attack otherwise the social welfare decreases. It means he system does not favour when attacker attacks. In subplot $(b)$ most of the games are at zero or around zero social welfare. It suggests, in many cases attack and defence are not favoured. Negative social welfare values indicates the scenarios where cost outweighs the benefits. Positive social welfare value means the system achieves beneficial outcomes.
\begin{figure}[H]
\centering
\begin{subfigure}{0.32\textwidth}
\includegraphics[width=1\linewidth, height=4cm]{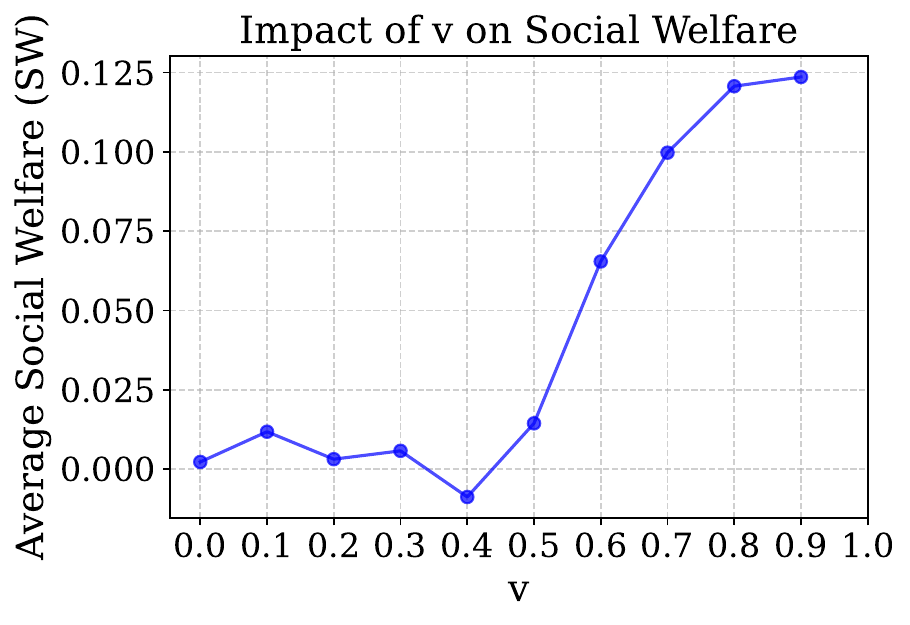}
\caption{}
\end{subfigure}
\begin{subfigure}{0.32\textwidth}
\includegraphics[width=1\linewidth, height=4cm]{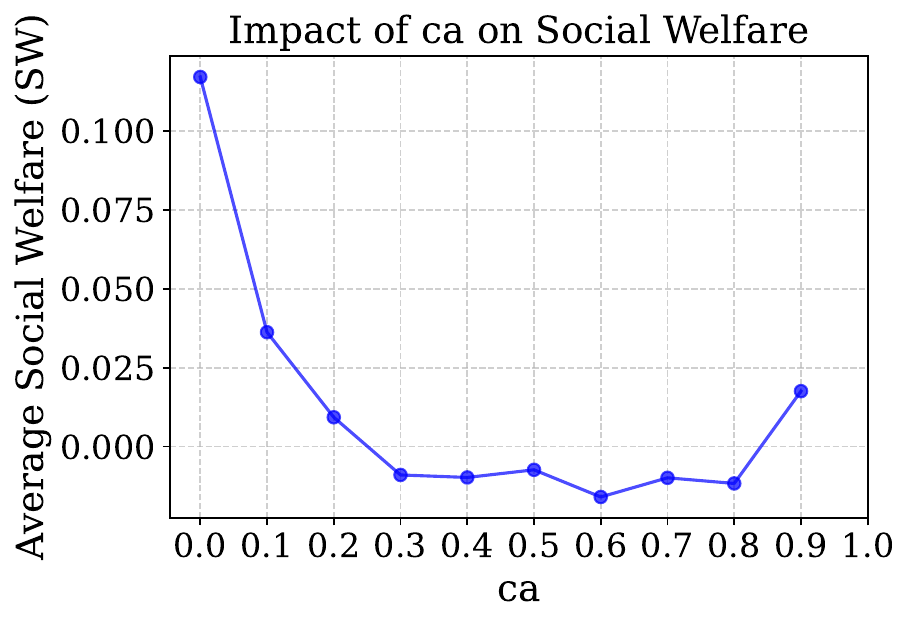}
\caption{}
\end{subfigure}
\begin{subfigure}{0.32\textwidth}
\includegraphics[width=1\linewidth, height=4cm]{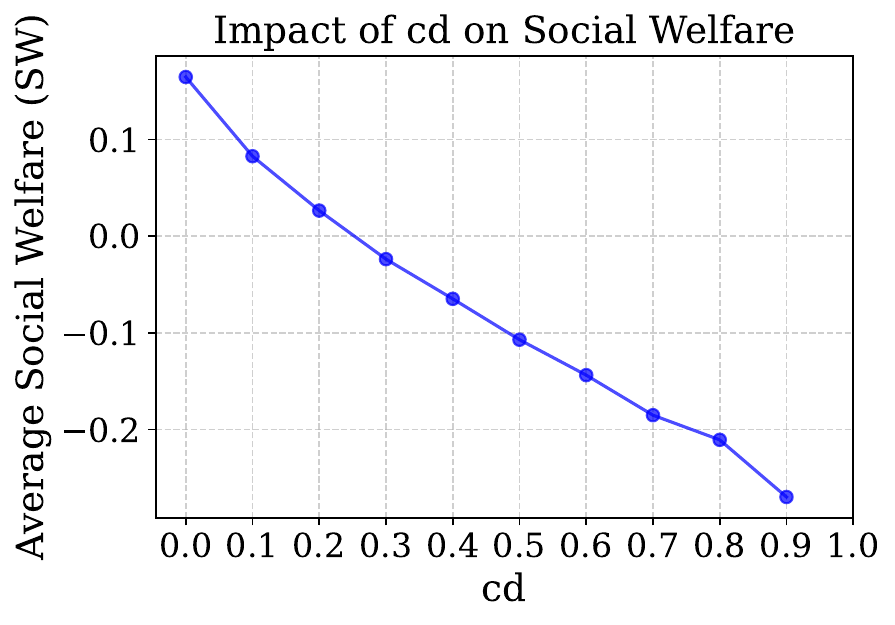}
\caption{}
\end{subfigure}
\caption{The impact of parameters on social welfare. Subplots $(a)$, $(b)$, and $(c)$ show a positive impact of parameters $v,c_a$ and $c_d$, respectively, on social welfare outcomes.}
\label{fig:18}
\end{figure}
The impact of different parameters on social welfare is analysed in Figure \ref{fig:18}, revealing key insights into cyber security decision-making. Subplot $(a)$, highlights that as intensity of defence $v$ increases, average social welfare also increases with a sharp rise beyond $v=0.5$. It indicates that low values of defence intensity do not favour the social welfare because they will attract more attacks. Subplot $(b)$ shows a sharp decline in social welfare as cost of attack $c_a$ increase from 0 to 0.2. It suggests that a slight increase in cost of attack discourages the attacker due to reduced benefit. The negative linear trend in the subplot $(c)$ indicates that social welfare steadily declines with increasing cost of defence because a high cost imposes a financial burden on defenders. Consequently, organizations may opt for minimal or no defence, increasing vulnerabilities, and reducing social welfare. Overall, these findings highlight the need of designing cost-effective security strategies that maximise protection. 

\section{Discussion and Conclusion}
This study highlights that cybersecurity is not a static process, but a dynamical one, where defence strategies must adapt to evolving threats. The use of Evolutionary Game Theory (EGT) to model cybersecurity system dynamics aligns with existing frameworks that analyse strategic interactions in complex systems, where one player's action affects another player's payoff \cite{Camerer+2004+374+392,sigmund2010calculus}. We modelled attacker-defender interactions as an asymmetric game, and used replicator dynamics to identify the equilibrium points and the stability conditions \cite{cressman2014replicator}. The stability analysis shows that in most of the random games, stability lies at the equilibrium point $(1,1)$ which means defenders always defend the systems and attackers continue attacks. However, catching the attackers and imposing penalties for the attacks can shift stability towards $(1,0)$ i.e., always defence and no attacks, consistent with real-world trends where robust defences deter attacks. For example, increasing the intensity of a successful defence $(v)$ significantly reduces attack attempts \cite{yang2021multi}. 

This work highlights the cost-benefit asymmetry: increasing defence intensity while lowering defence cost improves overall social welfare, while attackers are deterred by increasing the attack costs. Previous work \cite{tsen2022exploratory} has validated using real-world cyberattack data to confirm that higher defence investments correlate with a lower frequency of successful attack. Additionally, social welfare analysis highlights the economic impact of cybersecurity strategies. It is observed that increasing defence intensity while lowering defence cost improves overall social welfare. The findings show that robust defence mechanisms and strategic resource allocation are crucial for ensuring cybersecurity in dynamic digital environments. These findings also highlight the importance of collaborative efforts in the cybersecurity landscape, calling for industry-wide cooperation, and sharing of resources to boost collective defences \cite{solansky2021interorganizational}.

The existing literature on EGT analysis of cybersecurity and robust defence mechanisms is limited. More research on cyber security and threat prevention is needed as attacks are increasing in digital world. Our model focuses on one-on-one interactions, leaving multi-attacker scenarios or coordinated attacks as a promising direction for future research. It is challenging to accurately model the real world due to the limited availability of data for all attack types. We aim to  validate the findings from our model with real world data where appropriate (e.g. when relevant data is available). The integration of AI-driven systems, such as Machine Learning models, reinforcement learning agents, and large language models (LLMs), into this framework represents a promising path for future work. These advanced AI models can be trained to predict attacker behaviour, dynamically adapt defence strategies, and efficiently allocate resources, hence increasing resilience to sophisticated cyber threats and to ensure that defenders stay one step ahead in this evolving landscape. Another future direction highlighted by this research is the need of implications for cybersecurity policy and strategy. Small and Medium-sized
Enterprises (SMEs) need to focus on cost-effective defence strategies that enhance system stability without overwhelming resources. The findings support proactive measures, such as the creation of multi-layered defences and collaboration across industries to share intelligent and best practices.

To conclude, our EGT analysis not only sheds light on the fundamental dynamics of cyber interactions, but  also provides actionable insights for developing adaptive and robust systems. The findings show that robust defence mechanisms and strategic resource allocation are crucial for ensuring cybersecurity in dynamic digital environments. These findings also highlight the importance of collaborative efforts in the cybersecurity landscape, calling for industry-wide cooperation, and sharing of resources to boost collective defences. This cooperation will increase social welfare as an optimal balance of defence intensity and defence cost reduction leads to better security outcomes. 

\section*{Acknowledgements}
T.A.H. and Z.S. are supported by EPSRC (grant EP/Y00857X/1).

\bibliographystyle{unsrt}
\bibliography{mybib}

\begin{thebibliography}{10}

\bibitem{ahmad2024enhancing}
Basharat Ahmad, Zhaoliang Wu, Yongfeng Huang, and Sadaqat~Ur Rehman.
\newblock Enhancing the security in iot and iiot networks: An intrusion detection scheme leveraging deep transfer learning.
\newblock {\em Knowledge-Based Systems}, 305:112614, 2024.

\bibitem{sharif2022literature}
Md~Haris~Uddin Sharif and Mehmood~Ali Mohammed.
\newblock A literature review of financial losses statistics for cyber security and future trend.
\newblock {\em World Journal of Advanced Research and Reviews}, 15(1):138--156, 2022.

\bibitem{jang2014survey}
Julian Jang-Jaccard and Surya Nepal.
\newblock A survey of emerging threats in cybersecurity.
\newblock {\em Journal of computer and system sciences}, 80(5):973--993, 2014.

\bibitem{kumar2024effective}
Saurav Kumar and Ajit kumar Keshri.
\newblock An effective ddos attack mitigation strategy for iot using an optimization-based adaptive security model.
\newblock {\em Knowledge-Based Systems}, 299:112052, 2024.

\bibitem{7892931}
Ahmed~A. Alabdel~Abass, Liang Xiao, Narayan~B. Mandayam, and Zoran Gajic.
\newblock Evolutionary game theoretic analysis of advanced persistent threats against cloud storage.
\newblock {\em IEEE Access}, 5:8482--8491, 2017.

\bibitem{tariq2023critical}
Usman Tariq, Irfan Ahmed, Ali~Kashif Bashir, and Kamran Shaukat.
\newblock A critical cybersecurity analysis and future research directions for the internet of things: a comprehensive review.
\newblock {\em Sensors}, 23(8):4117, 2023.

\bibitem{o2018evolution}
Philip O'Kane, Sakir Sezer, and Domhnall Carlin.
\newblock Evolution of ransomware.
\newblock {\em Iet Networks}, 7(5):321--327, 2018.

\bibitem{aljaidi2022nhs}
Mohammad Aljaidi, Ayoub Alsarhan, Ghassan Samara, Raed Alazaidah, Sattam Almatarneh, Muhammad Khalid, and Yousef~Ali Al-Gumaei.
\newblock Nhs wannacry ransomware attack: Technical explanation of the vulnerability, exploitation, and countermeasures.
\newblock In {\em 2022 International Engineering Conference on Electrical, Energy, and Artificial Intelligence (EICEEAI)}, pages 1--6. IEEE, 2022.

\bibitem{akbanov2019wannacry}
Maxat Akbanov, Vassilios~G Vassilakis, and Michael~D Logothetis.
\newblock Wannacry ransomware: Analysis of infection, persistence, recovery prevention and propagation mechanisms.
\newblock {\em Journal of Telecommunications and Information Technology}, 75(1):113--124, 2019.

\bibitem{jia2023artificial}
Yan Jia, Zhaoquan Gu, Lei Du, Yu~Long, Ye~Wang, Jianxin Li, and Yanchun Zhang.
\newblock Artificial intelligence enabled cyber security defense for smart cities: A novel attack detection framework based on the mdata model.
\newblock {\em Knowledge-Based Systems}, 276:110781, 2023.

\bibitem{nguyen2024deep}
Van~Quan Nguyen, Long~Thanh Ngo, Viet~Hung Nguyen, Nathan Shone, et~al.
\newblock Deep clustering hierarchical latent representation for anomaly-based cyber-attack detection.
\newblock {\em Knowledge-Based Systems}, 301:112366, 2024.

\bibitem{rasikha2024ensemble}
V~Rasikha and P~Marikkannu.
\newblock An ensemble deep learning-based cyber attack detection system using optimization strategy.
\newblock {\em Knowledge-Based Systems}, 301:112211, 2024.

\bibitem{zhao2022cyber}
Jun Zhao, Minglai Shao, Hong Wang, Xiaomei Yu, Bo~Li, and Xudong Liu.
\newblock Cyber threat prediction using dynamic heterogeneous graph learning.
\newblock {\em Knowledge-Based Systems}, 240:108086, 2022.

\bibitem{yang2023highly}
Xingyuan Yang, Jie Yuan, Hao Yang, Ya~Kong, Hao Zhang, and Jinyu Zhao.
\newblock A highly interactive honeypot-based approach to network threat management.
\newblock {\em Future Internet}, 15(4):127, 2023.

\bibitem{petrosyan2023estimated}
Ani Petrosyan.
\newblock Estimated cost of cybercrime worldwide 2017-2028.
\newblock {\em Retrieved July}, 23:2023, 2023.

\bibitem{baruwal2024towards}
Mohan Baruwal~Chhetri, Shahroz Tariq, Ronal Singh, Fatemeh Jalalvand, Cecile Paris, and Surya Nepal.
\newblock Towards human-ai teaming to mitigate alert fatigue in security operations centres.
\newblock {\em ACM Transactions on Internet Technology}, 24(3):1--22, 2024.

\bibitem{juuso2013proactive}
Anna-Maija Juuso, Ari Takanen, and Kati Kittil{\"a}.
\newblock Proactive cyber defense: Understanding and testing for advanced persistent threats (apts).
\newblock In {\em 12th European Conference on Information Warfare and Security (ECIW)}, page 383, 2013.

\bibitem{heidt2019investigating}
Margareta Heidt, Jin~P Gerlach, and Peter Buxmann.
\newblock Investigating the security divide between sme and large companies: How sme characteristics influence organizational it security investments.
\newblock {\em Information Systems Frontiers}, 21:1285--1305, 2019.

\bibitem{9853515}
Alladean Chidukwani, Sebastian Zander, and Polychronis Koutsakis.
\newblock A survey on the cyber security of small-to-medium businesses: Challenges, research focus and recommendations.
\newblock {\em IEEE Access}, 10:85701--85719, 2022.

\bibitem{ilca2023enhancing}
Lucian~Florin Ilca, Ogru{\c{t}}an~Petre Lucian, and Titus~Constantin Balan.
\newblock Enhancing cyber-resilience for small and medium-sized organizations with prescriptive malware analysis, detection and response.
\newblock {\em Sensors}, 23(15):6757, 2023.

\bibitem{skopik2022detecting}
Florian Skopik, Markus Wurzenberger, and Max Landauer.
\newblock Detecting unknown cyber security attacks through system behavior analysis.
\newblock In {\em Cybersecurity of Digital Service Chains: Challenges, Methodologies, and Tools}, pages 103--119. Springer, 2022.

\bibitem{dorsey2003mathematical}
David~W Dorsey and Michael~D Coovert.
\newblock Mathematical modeling of decision making: a soft and fuzzy approach to capturing hard decisions.
\newblock {\em Human Factors}, 45(1):117--135, 2003.

\bibitem{smith1973logic}
J~Maynard Smith and George~R Price.
\newblock The logic of animal conflict.
\newblock {\em Nature}, 246(5427):15--18, 1973.

\bibitem{ohtsuki2008evolutionary}
Hisashi Ohtsuki and Martin~A Nowak.
\newblock Evolutionary stability on graphs.
\newblock {\em Journal of Theoretical Biology}, 251(4):698--707, 2008.

\bibitem{sigmund2010calculus}
Karl Sigmund.
\newblock {\em The calculus of selfishness}.
\newblock Princeton University Press, 2010.

\bibitem{khalid2023recent}
Mohd Nor~Akmal Khalid, Amjed~Ahmed Al-Kadhimi, and Manmeet~Mahinderjit Singh.
\newblock Recent developments in game-theory approaches for the detection and defense against advanced persistent threats (apts): a systematic review.
\newblock {\em Mathematics}, 11(6):1353, 2023.

\bibitem{SMITH198643}
John~Maynard Smith.
\newblock Evolutionary game theory.
\newblock {\em Physica D: Nonlinear Phenomena}, 22(1):43--49, 1986.
\newblock Proceedings of the Fifth Annual International Conference.

\bibitem{hofbauer1998evolutionary}
Josef Hofbauer and Karl Sigmund.
\newblock {\em Evolutionary games and population dynamics}.
\newblock Cambridge university press, 1998.

\bibitem{anderson2006economics}
Ross Anderson and Tyler Moore.
\newblock The economics of information security.
\newblock {\em science}, 314(5799):610--613, 2006.

\bibitem{patterson2019behavioral}
Wayne Patterson and Cynthia~E Winston-Proctor.
\newblock {\em Behavioral cybersecurity: Applications of personality psychology and computer science}.
\newblock CRC Press, 2019.

\bibitem{von2013information}
Rossouw Von~Solms and Johan Van~Niekerk.
\newblock From information security to cyber security.
\newblock {\em computers \& security}, 38:97--102, 2013.

\bibitem{tirenin1999concept}
Walt Tirenin and Don Faatz.
\newblock A concept for strategic cyber defense.
\newblock In {\em MILCOM 1999. IEEE Military Communications. Conference Proceedings (Cat. No. 99CH36341)}, volume~1, pages 458--463. IEEE, 1999.

\bibitem{hilbe2023evolutionary}
Christian Hilbe, Maria Kleshnina, and Kate{\v{r}}ina Sta{\v{n}}kov{\'a}.
\newblock Evolutionary games and applications: Fifty years of ‘the logic of animal conflict’.
\newblock {\em Dynamic Games and Applications}, 13(4):1035--1048, 2023.

\bibitem{Han2022emergent}
The~Anh Han.
\newblock Emergent behaviours in multi-agent systems with evolutionary game theory.
\newblock {\em AI Communications}, 35(4), 2022.

\bibitem{perc2017statistical}
Matja{\v{z}} Perc, Jillian~J Jordan, David~G Rand, Zhen Wang, Stefano Boccaletti, and Attila Szolnoki.
\newblock Statistical physics of human cooperation.
\newblock {\em Physics Reports}, 687:1--51, 2017.

\bibitem{traulsen2023future}
Arne Traulsen and Nikoleta~E Glynatsi.
\newblock The future of theoretical evolutionary game theory.
\newblock {\em Philosophical Transactions of the Royal Society B}, 378(1876):20210508, 2023.

\bibitem{cimpeanu2021cost}
Theodor Cimpeanu, Cedric Perret, and The~Anh Han.
\newblock Cost-efficient interventions for promoting fairness in the ultimatum game.
\newblock {\em Knowledge-Based Systems}, 233:107545, 2021.

\bibitem{liu2020evolutionary}
Linjie Liu and Xiaojie Chen.
\newblock Evolutionary game dynamics in multiagent systems with prosocial and antisocial exclusion strategies.
\newblock {\em Knowledge-Based Systems}, 188:104835, 2020.

\bibitem{JIA2024111962}
Danyang Jia, Chen Shen, Xiangfeng Dai, Xinyu Wang, Junliang Xing, Pin Tao, Yuanchun Shi, and Zhen Wang.
\newblock Freedom of choice disrupts cyclic dominance but maintains cooperation in voluntary prisoner’s dilemma game.
\newblock {\em Knowledge-Based Systems}, 299:111962, 2024.

\bibitem{ZHU2025113153}
Congcong Zhu, Dayong Ye, Tianqing Zhu, and Wanlei Zhou.
\newblock The evolution of cooperation in continuous dilemmas via multi-agent reinforcement learning.
\newblock {\em Knowledge-Based Systems}, 315:113153, 2025.

\bibitem{zhu2018evolutionary}
Guang Zhu, Hu~Liu, and Mining Feng.
\newblock An evolutionary game-theoretic approach for assessing privacy protection in mhealth systems.
\newblock {\em International journal of environmental research and public health}, 15(10):2196, 2018.

\bibitem{alalawi2019pathways}
Zainab Alalawi, The~Anh Han, Yifeng Zeng, and Aiman Elragig.
\newblock Pathways to good healthcare services and patient satisfaction: An evolutionary game theoretical approach.
\newblock In {\em Artificial Life Conference Proceedings}, pages 135--142. MIT Press One Rogers Street, Cambridge, MA 02142-1209, USA journals-info~…, 2019.

\bibitem{cimpeanu2022artificial}
Theodor Cimpeanu, Francisco~C Santos, Lu{\'\i}s~Moniz Pereira, Tom Lenaerts, and The~Anh Han.
\newblock Artificial intelligence development races in heterogeneous settings.
\newblock {\em Scientific Reports}, 12(1):1723, 2022.

\bibitem{han2020regulate}
Han The~Anh, Luis~Moniz Pereira, Francisco~C Santos, Tom Lenaerts, et~al.
\newblock To regulate or not: a social dynamics analysis of an idealised ai race.
\newblock {\em Journal of Artificial Intelligence Research}, 69:881--921, 2020.

\bibitem{milinski2008collective}
Manfred Milinski, Ralf~D Sommerfeld, Hans-J{\"u}rgen Krambeck, Floyd~A Reed, and Jochem Marotzke.
\newblock The collective-risk social dilemma and the prevention of simulated dangerous climate change.
\newblock {\em Proceedings of the National Academy of Sciences}, 105(7):2291--2294, 2008.

\bibitem{santos2011risk}
Francisco~C Santos and Jorge~M Pacheco.
\newblock Risk of collective failure provides an escape from the tragedy of the commons.
\newblock {\em Proceedings of the National Academy of Sciences}, 108(26):10421--10425, 2011.

\bibitem{pascoe2023public}
Cherilyn~E Pascoe.
\newblock Public draft: The nist cybersecurity framework 2.0.
\newblock {\em National Institute of Standards and Technology}, 2023.

\bibitem{fan2022survey}
Jiaxin Fan, Qi~Yan, Mohan Li, Guanqun Qu, and Yang Xiao.
\newblock A survey on data poisoning attacks and defenses.
\newblock In {\em 2022 7th IEEE International Conference on Data Science in Cyberspace (DSC)}, pages 48--55. IEEE, 2022.

\bibitem{etesami2019dynamic}
S~Rasoul Etesami and Tamer Ba{\c{s}}ar.
\newblock Dynamic games in cyber-physical security: An overview.
\newblock {\em Dynamic Games and Applications}, 9(4):884--913, 2019.

\bibitem{miao2018hybrid}
Fei Miao, Quanyan Zhu, Miroslav Pajic, and George~J Pappas.
\newblock A hybrid stochastic game for secure control of cyber-physical systems.
\newblock {\em Automatica}, 93:55--63, 2018.

\bibitem{zhang2023security}
Hengwei Zhang, Yan Mi, Yumeng Fu, Xiaohu Liu, Yuchen Zhang, Jindong Wang, and Jinglei Tan.
\newblock Security defense decision method based on potential differential game for complex networks.
\newblock {\em Computers \& Security}, 129:103187, 2023.

\bibitem{boudko2019adaptive}
Svetlana Boudko and Habtamu Abie.
\newblock Adaptive cybersecurity framework for healthcare internet of things.
\newblock In {\em 2019 13th International Symposium on Medical Information and Communication Technology (ISMICT)}, pages 1--6. IEEE, 2019.

\bibitem{zhang2021bayesian}
Yunxiao Zhang and Pasquale Malacaria.
\newblock Bayesian stackelberg games for cyber-security decision support.
\newblock {\em Decision Support Systems}, 148:113599, 2021.

\bibitem{yang2021multi}
Pengxi Yang, Fei Gao, and Hua Zhang.
\newblock Multi-player evolutionary game of network attack and defense based on system dynamics.
\newblock {\em Mathematics}, 9(23):3014, 2021.

\bibitem{hu2020optimal}
Hao Hu, Yuling Liu, Chen Chen, Hongqi Zhang, and Yi~Liu.
\newblock Optimal decision making approach for cyber security defense using evolutionary game.
\newblock {\em IEEE Transactions on Network and Service Management}, 17(3):1683--1700, 2020.

\bibitem{xu2020study}
Xiaotong Xu, Gaocai Wang, Jintian Hu, and Yuting Lu.
\newblock Study on stochastic differential game model in network attack and defense.
\newblock {\em Security and Communication Networks}, 2020(1):3417039, 2020.

\bibitem{zhang2018attack}
Hengwei Zhang, LV~Jiang, Shirui Huang, Jindong Wang, and Yuchen Zhang.
\newblock Attack-defense differential game model for network defense strategy selection.
\newblock {\em IEEE Access}, 7:50618--50629, 2018.

\bibitem{app13074645}
N’guessan Yves-Roland Douha, Masahiro Sasabe, Yuzo Taenaka, and Youki Kadobayashi.
\newblock An evolutionary game theoretic analysis of cybersecurity investment strategies for smart-home users against cyberattacks.
\newblock {\em Applied Sciences}, 13(7), 2023.

\bibitem{TOSH201827}
Deepak Tosh, Shamik Sengupta, Charles~A. Kamhoua, and Kevin~A. Kwiat.
\newblock Establishing evolutionary game models for cyber security information exchange (cybex).
\newblock {\em Journal of Computer and System Sciences}, 98:27--52, 2018.

\bibitem{schuster1983replicator}
Peter Schuster and Karl Sigmund.
\newblock Replicator dynamics.
\newblock {\em Journal of theoretical biology}, 100(3):533--538, 1983.

\bibitem{roca2009evolutionary}
Carlos~P Roca, Jos{\'e}~A Cuesta, and Angel S{\'a}nchez.
\newblock Evolutionary game theory: Temporal and spatial effects beyond replicator dynamics.
\newblock {\em Physics of life reviews}, 6(4):208--249, 2009.

\bibitem{yang2017basic}
King-Hay Yang.
\newblock {\em Basic finite element method as applied to injury biomechanics}.
\newblock Academic Press, 2017.

\bibitem{gandolfo1997economic}
Giancarlo Gandolfo.
\newblock {\em Economic dynamics: study edition}.
\newblock Springer Science \& Business Media, 1997.

\bibitem{golub2000eigenvalue}
Gene~H Golub and Henk~A Van~der Vorst.
\newblock Eigenvalue computation in the 20th century.
\newblock {\em Journal of Computational and Applied Mathematics}, 123(1-2):35--65, 2000.

\bibitem{taras1995solution}
AM~Taras'~yev.
\newblock The solution of evolutionary games using the theory of hamilton-jacobi equations.
\newblock {\em Journal of Applied Mathematics and Mechanics}, 59(6):921--933, 1995.

\bibitem{wall1998catching}
David~S Wall.
\newblock Catching cybercriminals: policing the internet.
\newblock {\em International Review of Law, Computers \& Technology}, 12(2):201--218, 1998.

\bibitem{duong2025evolutionary}
Manh~Hong Duong et~al.
\newblock Evolutionary dynamics with random payoff matrices.
\newblock {\em Europhysics Letters}, 149(3):32001, 2025.

\bibitem{han2012equilibrium}
The~Anh Han, Arne Traulsen, and Chaitanya~S Gokhale.
\newblock On equilibrium properties of evolutionary multi-player games with random payoff matrices.
\newblock {\em Theoretical Population Biology}, 81(4):264--272, 2012.

\bibitem{rand2011evolution}
David~G Rand and Martin~A Nowak.
\newblock The evolution of antisocial punishment in optional public goods games.
\newblock {\em Nature communications}, 2(1):434, 2011.

\bibitem{han2016emergence}
The~Anh Han.
\newblock Emergence of social punishment and cooperation through prior commitments.
\newblock In {\em Proceedings of the thirtieth aaai conference on artificial intelligence}, pages 2494--2500, 2016.

\bibitem{galla2013complex}
Tobias Galla and J~Doyne Farmer.
\newblock Complex dynamics in learning complicated games.
\newblock {\em Proceedings of the National Academy of Sciences}, 110(4):1232--1236, 2013.

\bibitem{duong2016analysis}
Manh~Hong Duong and The~Anh Han.
\newblock Analysis of the expected density of internal equilibria in random evolutionary multi-player multi-strategy games.
\newblock {\em Journal of Mathematical Biology}, 73:1727--1760, 2016.

\bibitem{broom1997multi}
M~Broom, C~Cannings, and GT~Vickers.
\newblock Multi-player matrix games.
\newblock {\em Bulletin of mathematical biology}, 59:931--952, 1997.

\bibitem{tsen2022exploratory}
Elinor Tsen, Ryan~KL Ko, and Sergeja Slapnicar.
\newblock An exploratory study of organizational cyber resilience, its precursors and outcomes.
\newblock {\em Journal of Organizational Computing and Electronic Commerce}, 32(2):153--174, 2022.

\bibitem{sangani2012cyber}
Nilaykumar~Kiran Sangani and Balakrishnan Vijayakumar.
\newblock Cyber security scenarios and control for small and medium enterprises.
\newblock {\em Informatica Economica}, 16(2):58, 2012.

\bibitem{oh2014need}
Sangkyo Oh and Kyungho Lee.
\newblock The need for specific penalties for hacking in criminal law.
\newblock {\em The Scientific World Journal}, 2014(1):736738, 2014.

\bibitem{tran2018law}
Delbert Tran.
\newblock The law of attribution: Rules for attribution the source of a cyber-attack.
\newblock {\em Yale JL \& Tech.}, 20:376, 2018.

\bibitem{khadam2023punish}
Nadia Khadam, Nasreen Anjum, Abu Alam, Qublai~Ali Mirza, Muhammad Assam, Emad~AA Ismail, and Mohamed~R Abonazel.
\newblock How to punish cyber criminals: A study to investigate the target and consequence based punishments for malware attacks in uk, usa, china, ethiopia \& pakistan.
\newblock {\em Heliyon}, 9(12), 2023.

\bibitem{8422719}
Afraa Attiah, Mainak Chatterjee, and Cliff~C. Zou.
\newblock A game theoretic approach to model cyber attack and defense strategies.
\newblock In {\em 2018 IEEE International Conference on Communications (ICC)}, pages 1--7, 2018.

\bibitem{sokri2018optimal}
Abderrahmane Sokri.
\newblock Optimal resource allocation in cyber-security: A game theoretic approach.
\newblock {\em Procedia computer science}, 134:283--288, 2018.

\bibitem{bada2019developing}
Maria Bada and Jason~RC Nurse.
\newblock Developing cybersecurity education and awareness programmes for small-and medium-sized enterprises (smes).
\newblock {\em Information \& Computer Security}, 27(3):393--410, 2019.

\bibitem{10.1145/3375627.3375829}
Alan Davoust and Michael Rovatsos.
\newblock Social contracts for non-cooperative games.
\newblock In {\em Proceedings of the AAAI/ACM Conference on AI, Ethics, and Society}, AIES '20, page 43–49, New York, NY, USA, 2020. Association for Computing Machinery.

\bibitem{han2024evolutionary}
The~Anh Han, Manh~Hong Duong, and Matjaz Perc.
\newblock Evolutionary mechanisms that promote cooperation may not promote social welfare.
\newblock {\em Journal of the Royal Society Interface}, 21(220):20240547, 2024.

\bibitem{apt2014selfishness}
Krzysztof~R Apt and Guido Sch{\"a}fer.
\newblock Selfishness level of strategic games.
\newblock {\em Journal of Artificial Intelligence Research}, 49:207--240, 2014.

\bibitem{Camerer+2004+374+392}
Colin~F. Camerer.
\newblock {\em CHAPTER THIRTEEN. Behavioral Game Theory: Predicting Human Behavior in Strategic Situations}, pages 374--392.
\newblock Princeton University Press, Princeton, 2004.

\bibitem{cressman2014replicator}
Ross Cressman and Yi~Tao.
\newblock The replicator equation and other game dynamics.
\newblock {\em Proceedings of the National Academy of Sciences}, 111(supplement\_3):10810--10817, 2014.

\bibitem{solansky2021interorganizational}
Stephanie~T Solansky and Tammy Beck.
\newblock Interorganizational information sharing: Collaboration during cybersecurity threats.
\newblock {\em Public Administration Quarterly}, 45(1):105--122, 2021.

\end{thebibliography}
\end{document}